\documentclass[a4paper,onecolumn,11pt,accepted=2025-05-17]{quantumarticle}
\pdfoutput=1
\usepackage[numbers]{natbib}
\usepackage[utf8]{inputenc}
\usepackage[english]{babel}
\usepackage[T1]{fontenc}

\usepackage[compatibility=false]{caption}
\usepackage{algorithm}
\usepackage{algpseudocode}
\usepackage{amsfonts}
\usepackage{amsmath}
\usepackage{amssymb}
\usepackage{arydshln}
\usepackage{authblk}
\usepackage{booktabs}
\usepackage{braket}
\usepackage{graphicx}
\usepackage{multirow}
\usepackage{xcolor}
\usepackage[colorlinks=true,citecolor=blue]{hyperref}
\usepackage{graphicx}
\DeclareMathOperator{\diag}{\rm diag}
\newcommand{\vect}[1]{\mathbf{#1}}

\title{Quantum DeepONet: Neural operators accelerated by quantum computing}

\author[1]{Pengpeng Xiao}
\orcid{0009-0003-8237-7847}
\affiliation{Department of Statistics and Data Science, Yale University, New Haven, CT 06511, USA}

\author[3]{Muqing Zheng}
\orcid{0000-0002-6659-9672}
\affiliation{Department of Industrial and Systems Engineering, Lehigh University, Bethlehem, PA 18015, USA}
\author[1]{Anran Jiao}
\orcid{0009-0007-6084-110X}
\affiliation{Department of Statistics and Data Science, Yale University, New Haven, CT 06511, USA}
\author[3]{Xiu Yang}
\orcid{0000-0003-0882-2650}
\email{ xiy518@lehigh.edu}
\affiliation{Department of Industrial and Systems Engineering, Lehigh University, Bethlehem, PA 18015, USA}
\author[1,4,*]{Lu Lu}
\orcid{0000-0002-5476-5768}
\email{lu.lu@yale.edu}
\affiliation{Department of Statistics and Data Science, Yale University, New Haven, CT 06511, USA}
\affiliation{Wu Tsai Institute, Yale University, New Haven, CT 06510, USA}

\date{}

\begin{document}
\maketitle

\begin{abstract}
In the realm of computational science and engineering, constructing models that reflect real-world phenomena requires solving partial differential equations (PDEs) with different conditions. Recent advancements in neural operators, such as deep operator network (DeepONet), which learn mappings between infinite-dimensional function spaces, promise efficient computation of PDE solutions for a new condition in a single forward pass. 
However, classical DeepONet entails quadratic complexity concerning input dimensions during evaluation. 
Given the progress in quantum algorithms and hardware, here we propose to utilize quantum computing to accelerate DeepONet evaluations, yielding complexity that is linear in input dimensions.
Our proposed quantum DeepONet integrates unary encoding and orthogonal quantum layers. We benchmark our quantum DeepONet using a variety of PDEs, including the antiderivative operator, advection equation, and Burgers' equation. We demonstrate the method's efficacy in both ideal and noisy conditions. Furthermore, we show that our quantum DeepONet can also be informed by physics, minimizing its reliance on extensive data collection. Quantum DeepONet will be particularly advantageous in applications in outer loop problems which require exploring parameter space and solving the corresponding PDEs, such as uncertainty quantification and optimal experimental design.
\end{abstract}

\section{Introduction}

Partial differential equations (PDEs) play a crucial role in modeling complex phenomena that are fundamental to both natural and engineered systems. Traditional numerical methods, such as finite difference, finite element, and finite volume methods, typically involve discretizing the solution space and solving finite-dimensional problems. These approaches, however, are computationally intensive and require a complete re-solving of equations with even minor adjustments to the system. Recently, neural networks have been employed to learn the solutions of PDEs~\cite{karniadakis2021physics, guo2016convolutional, yu2018deep, zhu2018bayesian, bhatnagar2019prediction, kochkov2021machine}. In particular, physics-informed neural networks (PINNs) embed the PDE residual into the loss term~\cite{karniadakis2021physics, raissi2019physics, lu2021deepxde}, demonstrating potential in solving both forward and inverse problems~\cite{chen2020physics, daneker2024transfer, daneker2023systems, fan2023deep}. Despite their promise, many of these methods remain mesh-dependent or require re-training when new functional parameters are introduced.

To address these limitations, deep neural operators have gained popularity for learning the mapping between infinite-dimensional spaces of functions through data~\cite{ lu2021learning, li2021fourier, anandkumar2020neural, li2023transformer, hao2023gnot, kovachki2023neural}. Once trained, neural operators are able to efficiently evaluate the PDE solutions for a new PDE instance in a single forward pass. Additionally, the output of neural operators can be discretized at different levels of resolutions or evaluated at any point. The training of neural operators can also incorporate physics priors~\cite{wang2021learning, li2024physics}, aligning the concept of PINNs, which has been shown to enhance accuracy significantly. The main categories of neural operators include integral kernel operators~\cite{anandkumar2020neural, li2021fourier, li2023fourier}, transformer-based neural operators~\cite{hao2023gnot, li2023transformer}, and DeepONet~\cite{lu2021learning}. Integral kernel operators, such as Fourier neural operator (FNO)~\cite{li2021fourier}, leverage iterative learnable kernel integration, but are usually restricted to grids. Transformer-based neural operator has larger model capacity but relies on sufficient data to achieve optimal performance. DeepONet, grounded in universal approximation theorem~\cite{chen1995universal}, on the other hand, can evaluate the solution of PDEs at any point in a mesh-free manner. There has been a wide range of developments of DeepONet~\cite{liu2021multiscale, jin2022mionet, zhu2023fourier, jiang2023fourier, mao2023ppdonet}, highlighting its adaptability in various complex systems. 

While classical developments greatly expand the potential of neural networks, quantum neural networks (QNNs) have also drawn much attention due to the potential of better complexity and higher capacities compared to their classical counterparts~\cite{biamonte2017quantum,cerezo2021variational, abbas2021power}.
Such advantages often directly come from the ability to efficiently encode and explore the exponentially large space on quantum computers~\cite{mitarai2018quantum}.
Specifically, there are quantum algorithms that demonstrate the quadratic speedup in online perceptron~\cite{kapoor2016quantum} and reinforcement learning~\cite{dunjko2016quantum}, as well as the exponential speedup in linear-system solving~\cite{PhysRevLett.103.150502, liu2024towards}, least-square fitting~\cite{wiebe2012quantum}, Boltzmann machine~\cite{amin2018quantum}, principal component
analysis~\cite{lloyd2014quantum}, and support vector machine~\cite{rebentrost2014quantum}.

Neural operators present an ideal application scenario for quantum neural networks designed to accelerate the evaluation process,
especially in situations where they are evaluated repeatedly in ``outer-loop problems'', such as forward uncertainty propagation and optimal experimental design. There is a recent development of quantum Fourier neural operator (QFNO)~\cite{jain2023quantum}. Utilizing a new form of the quantum Fourier transform, QFNO is expected to be substantially faster than classical FNO in evaluation. Instead of the linear number of evaluations required by classical FNO, QFNO only needs a logarithmic number of evaluations of the initial condition function, offering a significant improvement in efficiency. The success of QFNO motivates us to explore the possibility of accelerating other neural operators, such as DeepONet. 

However, as suggested by Refs.~\cite{wang2021noise, fontana2022non, thanasilp2023subtleties, schumann2024emergence, larocca2025barren}, the data embedding of classical datasets on quantum computers and hardware noise can induce barren plateaus and local minima that damage the trainability of quantum neural networks. This issue is especially problematic for optimizers that rely on the Fisher information matrix, as they require an exponentially large number of measurement shots to achieve accurate computation in barren plateaus~\cite{thanasilp2023subtleties}.

In this study, we design an architecture for quantum DeepONet and quantum physics-informed DeepONet (QPI-DeepONet).
To circumvent the trainability issue in QNN, we incorporate classical training and quantum evaluation by employing the orthogonal neural network structure outlined in Ref.~\cite{Landman2022quantummethods}. 
Our work preserves the quadratic speed-up with respect to the input dimension in the feed-forward pass from the quantum orthogonal neural network, with a minimal cost for classical data preprocessing before training. 
The results of our numerical experiments suggest the effectiveness of neural networks in solving different PDEs in both ideal and noisy environments. Furthermore, we conducted a detailed analysis of the impact of quantum noise on our quantum DeepONet, demonstrating how noise can influence performance and providing insights for the effect of noise mitigation strategies.

The paper is organized as follows. We first present the algorithm and architecture of quantum DeepONet in Section~\ref{sec: alg-qdeeponet}. In Section~\ref{sec: ideal_results}, we illustrate the ideal quantum simulation results of different applications of our quantum DeepONet. Then we investigate quantum noise and show the performance of the quantum DeepONet under two different noise models in Section~\ref{sec: noise_result}. Finally, we conclude our work and discuss the limitations in Section~\ref{sec: conclusion}. The background concepts related to quantum computing are provided in Appendices~\ref{sec: qcintro} and~\ref{app: density-matrix}. 
\section{Methods\label{sec: alg-qdeeponet}}

In this section, we first introduce a specific quantum circuit for network layers in Section~\ref{sec: quantum_layer}, referred to as ``quantum layers'', which are designed for constructing quantum orthogonal neural networks (Section~\ref{sec: QOrthoNN}). Building on these foundations, we propose a novel quantum DeepONet structure by synthesizing multiple quantum layers in Section~\ref{sec: quantum_deeponet}. The training method and loss function are detailed in Section~\ref{sec: training_QOrthoNN}. Furthermore, in addition to data-driven training, we also propose to use a physics-informed loss function, developing quantum physics-informed DeepONet (QPI-DeepONet) in Section~\ref{sec: QPI-DeepONet}. 


\subsection{Quantum methods for network layers \label{sec: quantum_layer}} 

\begin{figure}[htb]
\centering
\includegraphics[width=1\linewidth]{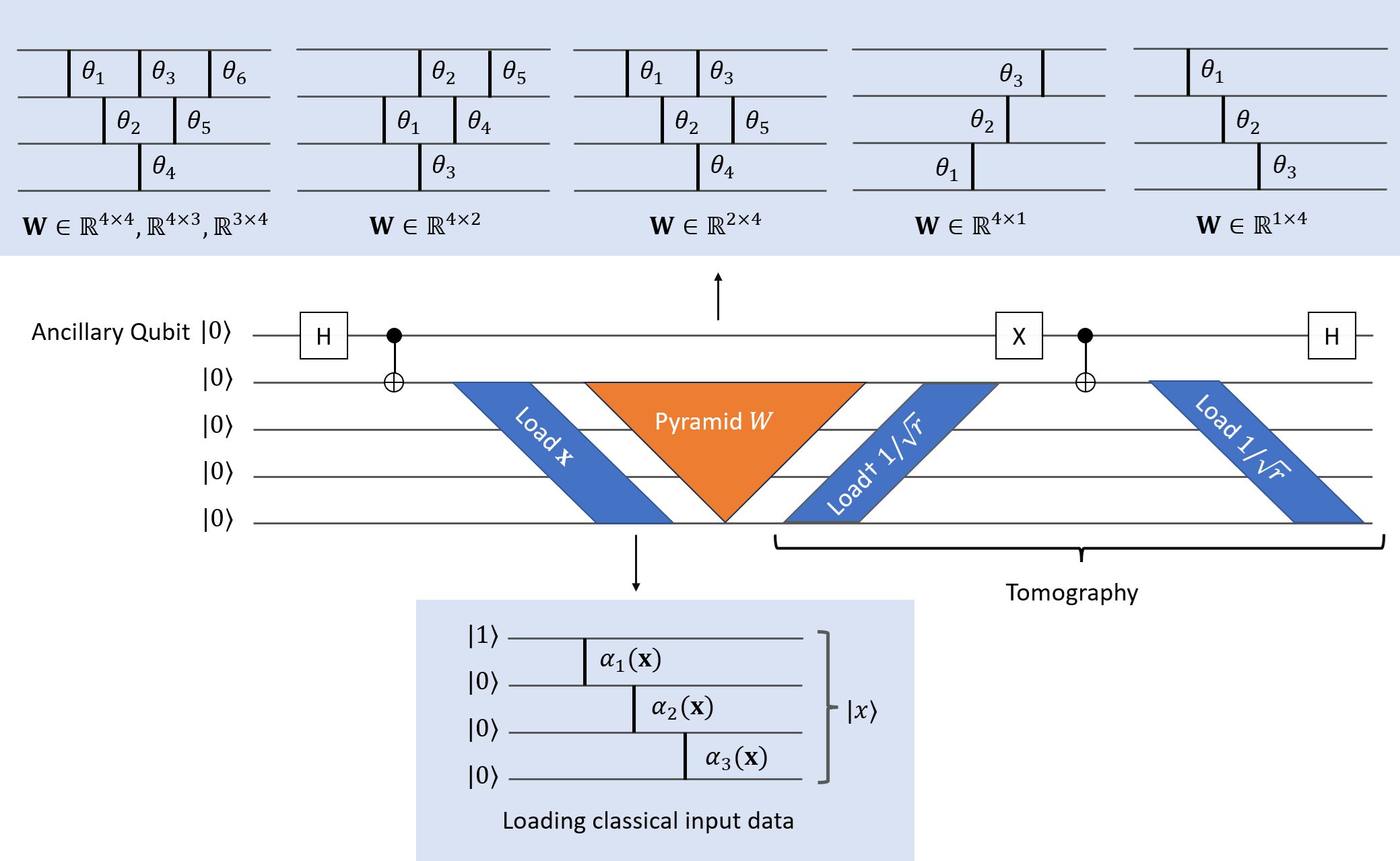}
\caption{\label{fig: quantum_layer}\textbf{The circuits of a quantum layer.} A quantum layer is composed of data loading, a pyramidal circuit, and tomography. An ancillary qubit is included for tomography. The vertical lines represent the two-qubit RBS gates, while the $\theta_i$ and $\alpha_i$ correspond to the parameter of the gate. We provide the example of a data loader for loading the classical vector $\vect{x} \in \mathbb{R}^{4}$ with $\Vert \vect{x} \Vert_2 = 1$. We demonstrate a quantum pyramidal circuit using all of the seven examples. $\vect{W} \in \mathbb{R}^{4\times 4}$, $\vect{W} \in \mathbb{R}^{4\times 3}$, and $\vect{W} \in \mathbb{R}^{3\times 4}$ share the same pyramidal circuit. The following circuit are other examples of $m\neq n$ cases: $\vect{W} \in \mathbb{R}^{4\times 2}$, $\vect{W} \in \mathbb{R}^{4\times 1}$, $\vect{W} \in \mathbb{R}^{2\times 4}$ and $\vect{W} \in \mathbb{R}^{1\times 4}$.}
\end{figure}

A classical neural network layer, with the input $\vect{x} \in \mathbb{R}^{n}$ and output $\vect{x'}\in\mathbb{R}^{m}$, takes the form $\vect{x'}=\sigma(\vect{W}\vect{x}+\vect{b})$. Here, $\vect{W} \in \mathbb{R}^{m\times n}$ represents the weight matrix, $\vect{b}\in \mathbb{R}^{m}$ is the bias, and $\sigma$ is the activation function. As demonstrated by Ref.~\cite{Landman2022quantummethods}, the matrix multiplication $\vect{W}\vect{x}$ can be accelerated by substituting the classical matrix multiplication with quantum matrix multiplication. The neural network layer accelerated by this quantum algorithm is referred as a quantum layer. We provide a detailed explanation of each step of a ``quantum layer'', beginning with an introduction to the basic gate, the reconfigurable beam splitter (RBS) gate, used in our method (Section~\ref{sec: RBS_gate}).
Because adding bias and applying non-linear transformation still require classical computation, the whole quantum layer, as depicted in Fig.~\ref{fig: quantum_layer}, involves three key steps to handle classical data on a quantum computer: (1) loading the classical data onto the quantum circuit (Section~\ref{sec: data_loading}), (2) performing matrix multiplication on quantum computer (Section~\ref{sec: pyramidal_circuit}), and (3) converting the resulting quantum data back into classical data (Section~\ref{sec: tomography}). In Section~\ref{sec:error-mitigation}, we illustrate a specific error mitigation method resulting from the unary encoding.
Finally, we summarize and provide the complexity of each step in Section~\ref{sec: quantum_layer_summary}.

\subsubsection{Reconfigurable beam splitter gate \label{sec: RBS_gate}} 

We first introduce reconfigurable beam splitter (RBS) gate~\cite{Landman2022quantummethods} as a basic tool used in our quantum layer:  
\begin{align*}
U_{RBS}(\theta)=
\begin{pmatrix} 
1 & 0 & 0 & 0\\ 
0 & \cos{\theta} & \sin{\theta} & 0\\
0 & -\sin{\theta} & \cos{\theta} & 0\\
0 & 0 & 0 & 1
\end{pmatrix},
\end{align*}
where its basis-gate decomposition is illustrated in Appendix~\ref{sec: qcintro}.
It performs rotation operation on state $\ket{01}\mapsto \cos{\theta} \ket{01}-\sin{\theta}\ket{10}$ and $\ket{10}\mapsto-\sin{\theta}\ket{01}+\cos{\theta}\ket{10}$, while leaving $\ket{00}$ and $\ket{11}$ unchanged. By carefully designing the circuit using RBS gates and setting $\theta$ to the required value, we can efficiently load data (Section~\ref{sec: data_loading}) and perform specialized matrix multiplication operations (Section~\ref{sec: pyramidal_circuit}). 

\subsubsection{Loading classical data input \label{sec: data_loading}} 

For a classical vector $\vect{x} \in \mathbb{R}^{n}$, to perform operations on quantum computers, this classical vector must be converted into a quantum state. This process is referred to as data loading. 

To load vector $\vect{x}$ onto a quantum state, it is crucial to ensure that the norm $\Vert \vect{x} \Vert_2 = 1$, as required by the probabilistic nature of quantum mechanics. If $\Vert \vect{x} \Vert_2 \neq 1$, normalization is required. We append an additional dimension to $\vect{x}$ at the first quantum layer in the neural network, which keeps the norm of $\vect{x}$ at 1 and in the meantime stores the information of the original norm of $\vect{x}$. Specifically, each element of $\vect{x}$ is first rescaled to the range $[-1,1]$. Then the value $\sqrt{1-\sum_{i} x_i^2/n}$ is assigned to the new dimension. This procedure can be viewed as data preprocessing before training, transforming the original $\vect{x}$ into
\begin{align*}
\begin{pmatrix} x_1 \\ x_2 \\ \ldots\\ x_n \\\sqrt{1-\sum_{i} x_i^2/n} \end{pmatrix},
\end{align*}
where $x_i$ is the $i$th element of $\vect{x}$. For subsequent quantum layers in the neural network, we simply divide $\vect{x}$ by $\Vert \vect{x} \Vert_2$ before loading the data.

The circuit for loading data is shown in Fig.~\ref{fig: quantum_layer} bottom. If $m=n$, the quantum circuit we adopt will have $n$ qubits, initialized such that the first qubit is at state $\ket{1}$ while the remaining qubits are $\ket{0}$. Then we apply a series of RBS gates parameterized by $(\alpha_{1},\alpha_2,\ldots,\alpha_{n-1})$, where
\begin{align*}
\alpha_1(\vect{x}) & = \arccos\left(x_1\right), \\
\alpha_2(\vect{x}) & = \arccos\left(x_2 \sin^{-1}(\alpha_1)\right), \\
\alpha_3(\vect{x}) & = \arccos\left(x_3 \sin^{-1}(\alpha_2) \sin^{-1}(\alpha_1)\right),
\end{align*} and so on. This sequence of operations converts the initial quantum state to 
\begin{align*}\ket{x}&=\cos{\alpha_1}\ket{10\ldots0}+\sin{\alpha_1}\cos{\alpha_2}\ket{01\ldots0}+\ldots+\sin{\alpha_1}\sin{\alpha_2}\ldots\sin{\alpha_{n-1}}\ket{00\ldots1}\\
&=x_{1}\underbrace{\ket{10\ldots0}}_{\ket{\vect{e}_1}}+x_2\underbrace{\ket{01\ldots0}}_{\ket{\vect{e}_2}}+\ldots+x_{n}\underbrace{\ket{00\ldots1}}_{\ket{\vect{e}_n}}.
\end{align*}
The configuration, where one qubit is in state $\ket{1}$, while all others are in state $\ket{0}$, is referred to as a ``unary state''. For simplicity, the $j$th unary state is denoted as $\ket{\vect{e}_j}$. Therefore, the information of $\vect{x}$ is encapsulated in $\ket{x}$ represented as the superposition of these unary states. Once data is loaded into the superposition of unary states, all of our subsequent operations, which utilize the RBS gate and only include transformations between $\ket{\vect{e}_j}$ states, are effectively confined to these unary states. This implies that the unary subspace throughout the entire process, allows us to employ the tomography, which will be further explained in Section~\ref{sec: tomography}.

On the other hand, when output and input dimensions are different $(m\neq n)$, the number of qubits required in the circuit will be $\max (m,n)$. If the output dimension is smaller than the input dimension, i.e., $m<n$, the data is loaded onto all $n$ qubits. Conversely, if $m>n$, the classical vector $\vect{x}$ is loaded onto the bottom $n$ qubits in the circuit, leaving upper $m-n$ qubits at $\ket{0}$.

\subsubsection{Quantum pyramidal circuit \label{sec: pyramidal_circuit}}

When classical data $\vect{x}$ is loaded onto the quantum circuit, matrix multiplication $\vect{y}=\vect{W}\vect{x}$, where $\vect{y}\in\mathbb{R}^{m}$, can be performed in quantum space. Here we adopt the quantum pyramidal circuit proposed in Ref.~\cite{Landman2022quantummethods}. Such pyramidal circuit features orthogonal matrix multiplication, meaning that the corresponding $\vect{W}$ is orthogonal.  

We first introduce the quantum pyramidal circuit for $m=n$ cases. The basic idea of this method is to decompose the orthogonal matrix $\vect{W}$ into a series of rotation matrices, which can be represented by RBS gates. These decomposed rotation matrices can be parameterized with angles ${\theta_{1},\theta_2\ldots,\theta_{d}}$, where $d = n(n-1)/2$. All of the parameterized RBS gates are arranged in a pyramid configuration. We take the $\vect{W}\in \mathbb{R}^{4\times 4}$ matrix in Fig.~\ref{fig: quantum_layer} as an example. 
On the loaded vector $\vect{x}$, the pyramidal portion of the circuit conducts $\vect{y} = \vect{W}\vect{x}$, where $\vect{W}$ is

{\tiny
\begin{align*}
\begin{pmatrix} 
C_{\theta_1} & S_{\theta_1} & & \\ 
-S_{\theta_1} & C_{\theta_1} & & \\
 &  & 1 &  \\
& & & 1 
\end{pmatrix}
\begin{pmatrix} 
1 & & & \\ 
 & C_{\theta_2} & S_{\theta_2}& \\
 & -S_{\theta_2} & C_{\theta_2} &  \\
& & & 1 
\end{pmatrix}
\begin{pmatrix} 
 C_{\theta_3} & S_{\theta_3} & & \\ 
 -S_{\theta_3} & C_{\theta_3} & & \\
 & & C_{\theta_4} & S_{\theta_4} \\
 & & -S_{\theta_4} & C_{\theta_4} 
\end{pmatrix}
\begin{pmatrix} 
1 & & & \\ 
 & C_{\theta_5} & S_{\theta_5}& \\
 & -S_{\theta_5} & C_{\theta_5} &  \\
& & & 1 
\end{pmatrix}
\begin{pmatrix} 
C_{\theta_{6}} & S_{\theta_{6}} & & \\ 
-S_{\theta_{6}} & C_{\theta_{6}} & & \\
 &  & 1 &  \\
& & & 1 
\end{pmatrix}.
\end{align*}}
$C_{\theta_j}$ and $S_{\theta_j}$ are $\cos \theta_j$ and $\sin \theta_j$ for any $j$, respectively.
Therefore, the resulting quantum state is
\begin{align*}
\ket{y}=\ket{Wx}=\sum_{i,j}W_{ji}x_i\ket{\vect{e}_j}.
\end{align*}

If $m\neq n$, the construction of the pyramidal circuit is the same as Ref.~\cite{Landman2022quantummethods}. Examples of this include $\vect{W} \in \mathbb{R}^{4\times 1}$, $\vect{W} \in \mathbb{R}^{1\times 4}$ and so on, as shown in Fig.~\ref{fig: quantum_layer}. Note that for $|m-n|=1$ cases, the pyramidal circuit is the same as $m=n$ cases. However, due to the difference in data loading and tomography process, the quantum layer is distinct. 

\subsubsection{Tomography for extracting classical output \label{sec: tomography}} After performing matrix multiplication in quantum space, it is necessary to convert the quantum information back to classical form for further processing, such as adding bias and applying non-linear transformation. This process is known as tomography. Tomography could commonly be expensive in terms of quantum resources when extracting the complete information from quantum states~\cite{aaronson2015read, huang2020predicting, van2023quantum}. However, in our method, the usage of unary state sparsely encodes information in Hilbert space and provides a feasible and efficient tomography method. This tomography method, proposed by Ref.~\cite{Landman2022quantummethods}, is illustrated in Fig.~\ref{fig: quantum_layer} middle.

We introduce an ancillary qubit and implement a Hadamard ($H$) and a CNOT gate between the ancillary qubit and the first data loader qubit before loading data (see Appendix~\ref{sec: qcintro} for the definition of gates). After the pyramid gate, the circuit performs an adjoint operation of the data loader of a uniform norm-1 vector $\left(\frac{1}{\sqrt{r}},\frac{1}{\sqrt{r}},\ldots,\frac{1}{\sqrt{r}}\right)$, where $r=\max (m,n)$ represents the number of qubits excluding the ancillary qubit. This is followed by an X gate and a CNOT gate. 

Finally, we load $\left(\frac{1}{\sqrt{r}},\frac{1}{\sqrt{r}},\ldots,\frac{1}{\sqrt{r}}\right)$  and a Hadamard gate. \textcolor{black}{In this way, the final quantum state in the circuit is
\begin{align}
    \frac{1}{2} \sum_{j} \left(  \sum_{i} W_{ji}x_{i} + \frac{1}{\sqrt{r}}  \right) \ket{0,\vect{e}_{j}} +  \frac{1}{2} \sum_{j} \left(  \sum_{i} W_{ji}x_{i} - \frac{1}{\sqrt{r}}  \right) \ket{1,\vect{e}_{j}}, \label{equ: tomo-state}
\end{align}
where $\ket{\xi,\vect{e}_{j}}$ indicates the ancillary qubit is in state $\ket{\xi}$ for $\xi \in \{0,1\}$, and the rest qubits are in the state $\ket{\vect{e}_{j}}$. Thus, for each $y_j = \sum_{i} W_{ji}x_{i}$, the tomography includes two steps:
\begin{enumerate}
    \item Sign recovery: noticing that 
    \begin{align*}
    {\rm Pr}[0,\vect{e}_{j}]-{\rm Pr}[1,\vect{e}_{j}] = 1/\sqrt{r}  \sum_{i} W_{ji}x_{i} = y_j/\sqrt{r},
    \end{align*}
    we have
    \begin{align*}
        {\rm sign}(y_j) = \begin{cases}
            +1, \text{ if } {\rm Pr}[0,\vect{e}_{j}] \geq {\rm Pr}[1,\vect{e}_{j}] \\
            -1, \text{ otherwise}
        \end{cases}.
    \end{align*}
    \item Value recovery: 
    \begin{align}
         y_j = \begin{cases}
              {\rm sign}(y_j) \cdot \left(2 \sqrt{{\rm Pr}[0,\vect{e}_{j}]}  - \frac{1}{\sqrt{r}} \right), \text{ if } {\rm sign}(y_j) > 0 \\
              {\rm sign}(y_j) \cdot \left(2 \sqrt{{\rm Pr}[1,\vect{e}_{j}]}  + \frac{1}{\sqrt{r}} \right), \text{ otherwise}
         \end{cases}\label{equ: tomography}.
    \end{align}
\end{enumerate}
In this scheme, the error of the estimating $y_j$ is independent of the dimension of the problem $r$. As the tomography circuit has depth $\mathcal{O}(n)$, the tomography process maintains a complexity of $\mathcal{O}(n/\delta^2)$ where $\delta$ is the tomography error and $1/\delta^2$ is the number of measurements. The scale of the tomography error is further explained in Section~\ref{sec: finite-sampling}.
}

When the output dimension is smaller than the input dimension ($m<n$), the tomography circuit is still the same, but only the information of bottom $m$ qubits are finally considered. In other words, the $\vect{e}_j$ in Eqs.~\eqref{equ: tomo-state} to \eqref{equ: tomography} refers to $j$th unary state for the bottom $m$ qubits. Consequently, the output $\sum_{i} W_{ji}x_{i}$ is restricted to size $m$. 

\subsubsection{Error mitigation\label{sec:error-mitigation}}

If the quantum circuit includes quantum noise, error mitigation methods could be introduced during tomography based on the system's properties. We apply the same method as in Ref.~\cite{Landman2022quantummethods}, where only unary measurement outcomes are kept and all the other non-unary outcomes are discarded. For example, if the measurement results in $0010$, it is a valid result to keep; if the measurement yields $0110$, then this result is ignored because it is not a unary string.
This post-selection technique is a benefit of unary encoding. The improvement on the $L^2$ error from this method, along with the percentages of kept and discarded shots, will be demonstrated in Figs.~\ref{fig: ODE_noise}B, C, and D in Section~\ref{sec: noise_ODE}.  In general, it provides a significant decrease in $L^2$ error across different levels of depolarizing error and finite-sampling error.

\subsubsection{Summary and remarks}\label{sec: quantum_layer_summary}
 
In conclusion, the structure of a complete quantum layer is shown in Fig.~\ref{fig: quantum_layer}. The number of qubits needed is $1+\max(m,n)$ for $\vect{W}\in \mathbb{R}^{m\times n}$, in which the top $1$ qubit is an ancillary qubit for tomography purposes, while other qubits are used to store information and perform operations.
Sequentially, we implement data loading, pyramidal circuit, and tomography, and thus complete the matrix multiplication in quantum space. 

\begin{figure}[htb]
    \centering
    \includegraphics[width=0.45\linewidth]{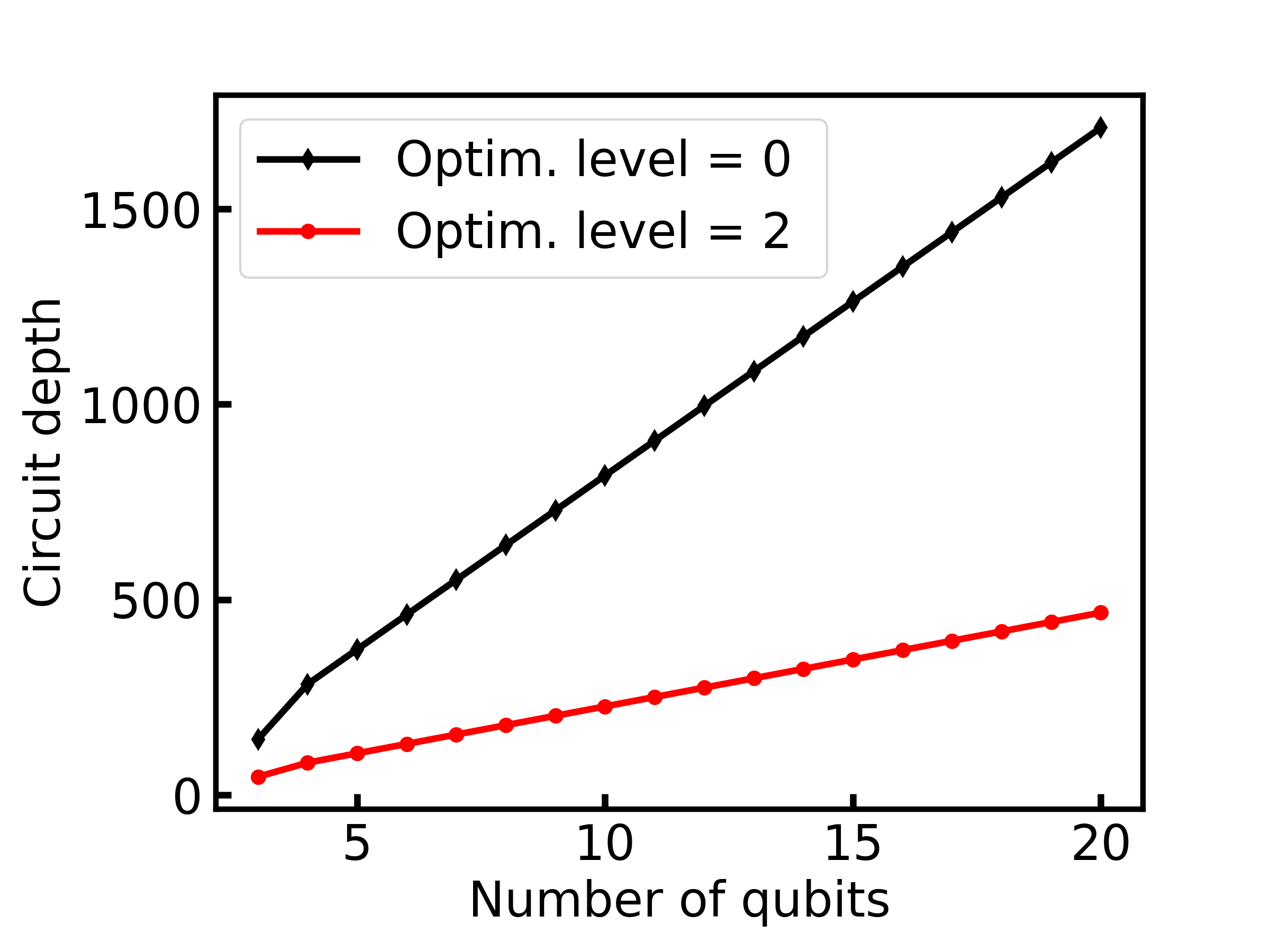}
    \caption{\textbf{Circuit depth growth for a circuit of the single all-to-all quantum layer in Fig.~\ref{fig: quantum_layer}.} The basis gate set is $\{ECR, RZ, SX, X\}$. Two levels of gate optimization with Qiskit transpiler are considered: level $0$ means no optimization and level $2$ includes qubit-layout optimization, inverse cancellation, 1-qubit gate optimization, and commutative cancellation.}
    \label{fig:depth-layer}
\end{figure}

\paragraph{Complexity.} 
Quantum layers can accelerate the feedforward pass, achieving a complexity of $O(n/\delta^2)$. Here, $\delta$ is the threshold for the tomography error. This complexity is the result of the full $\mathcal{O}(n)$-depth circuit in Fig.~\ref{fig: quantum_layer} with $\mathcal{O}(1/\delta^2)$ number of measurements in the data extraction. We will explicitly discuss the statistical error scaling in the data extraction in Section~\ref{sec: finite-sampling}. The complexities of all the components of quantum layers are shown in Table~\ref{tab: complexity_1}. The depth of a single-layer circuit in Fig.~\ref{fig: quantum_layer} has at most $3n+\mathcal{O}(1)$ RBS gates. This is counted by first considering the bigger pyramidal group of RBS gates with depth at most $2n+1$ by combining the component ``Load $\vect{x}$'', ``Pyramid $\mathbf{W}$'', and ``Load$^\dagger$ $1/\sqrt{r}$''. Then, the right-most ``Load $1/\sqrt{r}$'' has depth $n-1$ and there are some extra $\mathcal{O}(1)$ depths from the remaining gates related to the ancillary qubit. Providing the transpilation of the RBS gate in Appendix~\ref{sec: qcintro}, the circuit in Fig.~\ref{fig: quantum_layer} maintains an $\mathcal{O}(n)$ depth in terms of basis gates. Fig.~\ref{fig:depth-layer} illustrates this linear relation with and without gate optimization from Qiskit.
\begin{table}[htbp]
    \centering
    \caption{\textbf{Complexity of each step of a quantum layer.} Here, $n$ is the input dimension, and $\delta$ represents the threshold for the tomography error.\label{tab: complexity_1}}
   \begin{tabular}[t]{c|ccc}
    \toprule
    \centering
        Operation &  Doading input data & Quantum pyramidal circuit& Extracting output \\ \midrule
     Complexity    &  $\mathcal{O}(n)$ & $\mathcal{O}(n)$ & $\mathcal{O}(n/\delta^2)$ \\ \bottomrule
    \end{tabular}
\end{table}

\subsection{Quantum orthogonal neural network \label{sec: QOrthoNN}}
\begin{figure}[htb]
\centering
\includegraphics[width=0.9\linewidth]{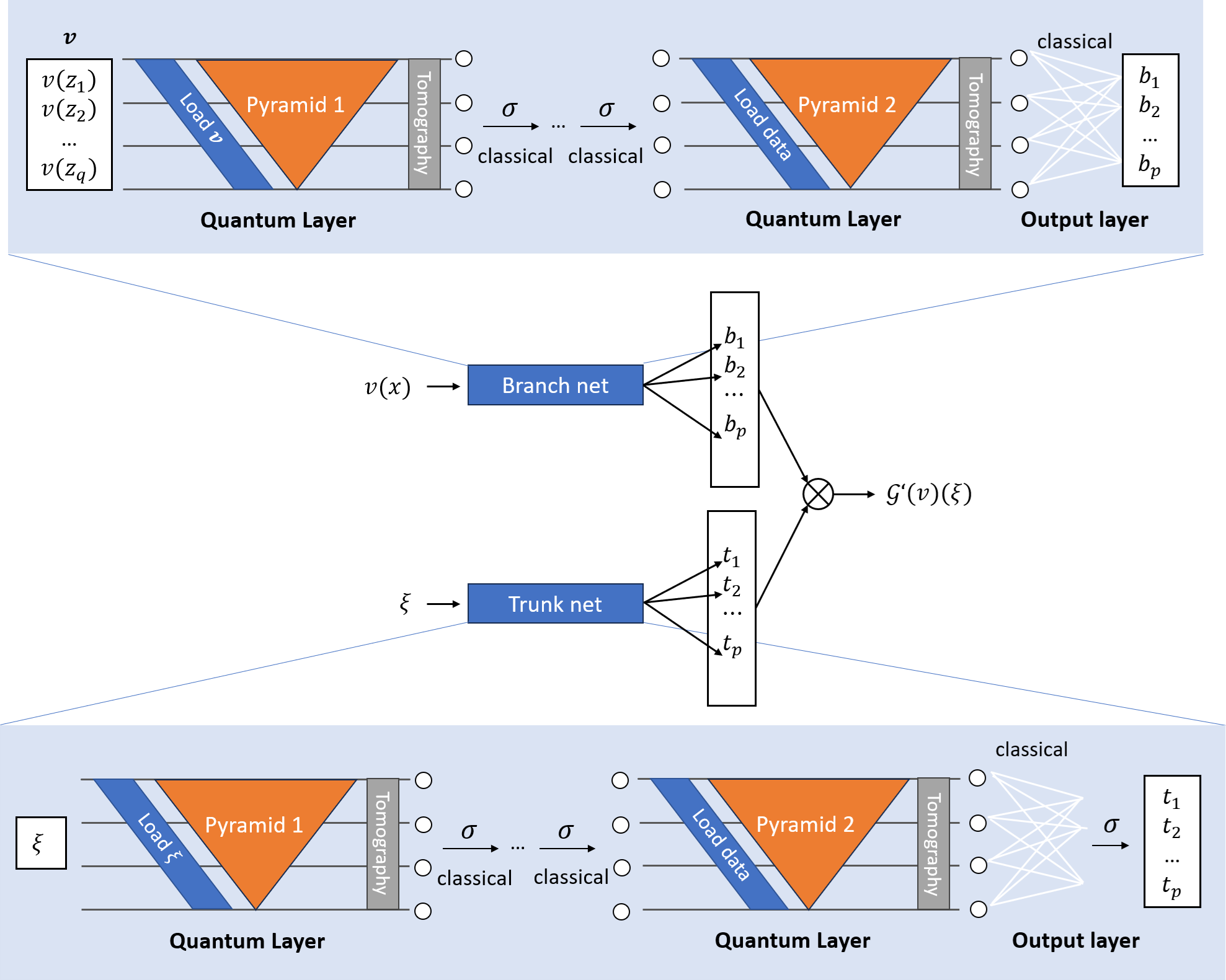}
\caption{\label{fig: quantum_deeponet}\textbf{Architecture of quantum DeepONet.} DeepONet consists of two subnetworks: the branch net and the trunk net. In quantum DeepONet, we replace these with QOrthoNN, which is composed of several quantum layers arranged sequentially. The addition of biases and nonlinear operations are performed on classical computers.}
\end{figure}

By integrating multiple quantum layers, we can construct a quantum orthogonal neural network (QOrthoNN). The input vector goes through a linear transformation in quantum space and is then measured and convert to classical space (Fig.~\ref{fig: quantum_deeponet}). Although not shown in the diagram, we add bias and apply non-linear transform thereafter classically. We proceed to the next layer and perform similar process. The sequence can be repeated several times until we reach the last layer, which consists solely of a classical linear transform. The dimension and norm of the quantum neural network output of is determined by the output layer, giving that former quantum layers always constrain the norm of processed vector to be 1. 

\subsection{Quantum DeepONet\label{sec: quantum_deeponet}}
DeepONet is a neural network architecture that aims to learn operators mapping between two infinite-dimensional function spaces. The most popular application of DeepONet is solving PDEs. Our goal is often to predict functions satisfying the PDEs under varying conditions, which could be the initial conditions, boundary conditions or coefficient fields of the PDEs. We define the input function $v\in\mathcal{V}$  over the domain $D \subset \mathbb{R}^{d}$ as 
\begin{align*}
v:D\ni x \mapsto v(x)\in \mathbb{R},
\end{align*}
and similarly, we define the output function $u\in\mathcal{U}$ over $D' \subset\mathbb{R}^{d'}$, which is described as
\begin{align*}
u:D' \ni \xi \mapsto u(\xi)\in\mathbb{R}.
\end{align*}
Suppose $\mathcal{V}$ and $\mathcal{U}$ are Banach spaces, and consider a parametric PDE taking the form
\begin{align*}
\mathcal{N}(v,u)=0,
\end{align*}
where $\mathcal{N}$ is a differential operator. The mapping between the input function space $\mathcal{V}$ and output function space $\mathcal{U}$ is defined the operator: 
\begin{align*}
\mathcal{G}: \mathcal{V}\ni v \mapsto u \in\mathcal{U}.
\end{align*}
DeepONet, therefore, is used to approximate $\mathcal{G}$.

A DeepONet includes a branch net and trunk net, each with an equivalent number of output neuron, denoted by $p$. The branch and trunk nets can adopt arbitrary architectures, like fully connected neural network (FNN), convolutional neural network (CNN), recurrent neural network (RNN), and residual neural network (ResNet). A diagrammatic representation of DeepONet is illustrated in the center of Fig.~\ref{fig: quantum_deeponet}. The branch network receives the input function evaluated at a discrete set of points $\{z_1,z_2,\ldots,z_q\}$, represented by $[v(z_1),v(z_2),\ldots,v(z_q)]$. The trunk net is fed with the location $\xi$ at which the output function is evaluated, which can include both time and space coordinates. The outputs of the branch and trunk networks are denoted by $[b_1(v),b_2(v),\ldots,b_p(v)]$ and $[t_1(\xi),t_2(\xi),\ldots,t_p(\xi)]$. Thus, the final output of DeepONet is the sum of the dot product of the branch and trunk network outputs and a bias $b_0\in\mathbb{R}$, expressed as 
\begin{align*}
\mathcal{G'}_{\theta}(v)(\xi)=\sum^{p}_{k=1}b_k(v)t_k(\xi)+b_0,
\end{align*}
where $\mathcal{G'}$ denotes the learned approximation of operator $\mathcal{G}$, and $\theta$ is the trainable parameter of the network. 

In this work, we propose a modification to the DeepONet framework by replacing the conventional branch and trunk networks with QOrthoNN (Fig.~\ref{fig: quantum_deeponet}). We refer to the resulting model as quantum DeepONet. 

\subsection{Training quantum DeepONet}\label{sec: training_QOrthoNN}

Up to this point, we have introduced QOrthoNN and the quantum DeepONet, but we have not yet discussed the training process of these quantum networks. Adapting the backpropagation scheme from Ref.~\cite{Landman2022quantummethods} for the pyramidal circuit, we train the network on classical computers, utilizing a classical orthogonal neural network (OrthoNN) that shares the same mathematical expression as QOthoNN. OrthoNN benefits from the properties of orthogonality, such as improved accuracy and better convergence during training~\cite{OrthNNBenefits, li2019orthogonal}, while maintaining the same asymptotic running time as a standard neural network. After training, we substitute the angular parameters of the RBS gates in the quantum circuits with trained paramters during the evaluation phase. It is during this evaluation phase on quantum computers that we anticipate significant acceleration benefits.

The comparison between the QOrthoNN, OrthoNN and the standard neural network is presented in Table~\ref{tab: complexity_2}. 
While OrthoNN and standard neural network both have a quadratic dependency on the input dimension $n$ for the forward pass, the QOrthoNN only requires a linear dependency, achieving a quadratic improvement in terms of the input dimension.
This reduction in computational complexity is especially advantageous in scenarios where the input dimension is large and frequent evaluations are needed, such as in the quantum DeepONet.

\begin{table}[htbp]
    \centering
    \caption{\textbf{Comparison of complexity for three networks.} $n$ and $\delta$ represent the input dimension and threshold for the tomography error, respectively.}\label{tab: complexity_2}
    {\footnotesize
   \begin{tabular}[t]{c|cc}
    \toprule
    \centering
        Algorithm & Feedforward pass & Weight matrix update \\ \midrule
        Quantum orthogonal neural network (QOrthoNN)~\cite{Landman2022quantummethods} & $\mathcal{O}(n/\delta^2)$ & -- \\
        \midrule
        Classical orthogonal neural network (OrthoNN)~\cite{Landman2022quantummethods} & $\mathcal{O}(n^2)$ & $\mathcal{O}(n^2)$ \\ 
        \midrule
        Standard neural network & $\mathcal{O}(n^2)$ & $\mathcal{O}(n^2)$ \\  \bottomrule
    \end{tabular}
    }
\end{table}

For data-driven training of quantum DeepONet, we sample $N$ distinct input functions $\{v^{(i)}\}^N_{i=1}$ from $\mathcal{V}$, and $Q$ locations $\{\xi^{(i)}_j\}^Q_{j=1}$ in the domain of ${G}(v^{(i)})$ for each input function $v^{(i)}$ as the inputs of training dataset. The corresponding solution $\mathcal{G}(v^{(i)})(\xi^{(i)}_j)$ is taken as the label of training dataset. The loss of DeepONet can therefore be expressed as
\begin{equation}
    \mathcal{L}_\text{operator}(\theta)=\frac{1}{NQ}\sum_{i=1}^{N}\sum_{j=1}^{Q}\left|\mathcal{G'}_{\theta}(v^{(i)})(\xi^{(i)}_j)-\mathcal{G}(v^{(i)})(\xi^{(i)}_j)\right|^2.
    \label{equ: operator_loss}
\end{equation}

To summarize, the workflow of our quantum method is divided into three distinct phases:
\begin{itemize}
    \item Training quantum DeepONet on classical computer;
    \item Transferring of parameters to quantum layer;
    \item Execution on quantum computer or simulator for evaluation.
\end{itemize}
Hence, the primary speedup offered by our method originates from the evaluation phase, while the training phase maintains a complexity comparable to that of standard neural networks.

\subsection{Quantum physics-informed DeepONet \label{sec: QPI-DeepONet}}
We further introduce physics-informed loss term during training,
\begin{equation}\mathcal{L}_{\text {physics}}(\boldsymbol{\theta})=\frac{1}{N Q} \sum_{i=1}^N \sum_{j=1}^Q\left|\mathcal{N}\left(v^{(i)}, \mathcal{G'}_{\theta}(v^{(i)})(\xi_j^{(i)})\right)\right|^2.\label{eq: physics loss}\end{equation}
The total loss function is therefore 
\begin{align*}
\mathcal{L}(\boldsymbol{\theta})=\mathcal{L}_{\text {physics}}(\boldsymbol{\theta})+\mathcal{L}_{\text {operator}}(\boldsymbol{\theta}),
\end{align*}
where $\mathcal{L}_{\text {operator}}$ has the same definition as Eq.~\eqref{equ: operator_loss}, but only includes the initial conditions and boundary conditions. By introducing the physics information into our network, we can reduce the demand of data and even train the network in the absence of solution input-output pairs. We name such architecture as quantum physics-informed DeepONet (QPI-DeepONet). In evaluation stage, QPI-DeepONet follows the same procedure as ordinary quantum DeepONet.

In some cases, we can embed boundary conditions into the network architecture, known as hard constrain~\cite{lu2022comprehensive}. For example, to enforce Dirichlet BCs $\mathcal{G}_{\theta}(v)(\xi)=g(\xi)$ for $\xi\in\varGamma_D$, we can construct the quantum DeepONet output as 
\begin{align*}
\mathcal{G''}_{\theta}(v)(\xi)=g(\xi)+\ell(\xi)\mathcal{G'}_{\theta}(v)(\xi),
\end{align*}
where $\mathcal{G'}_{\theta}(v)(\xi)$ is the output of vanilla quantum DeepONet, and $\ell(\xi)$ satisfy\[ \left\{ \begin{array}{l}
    \ell(\xi)=0,\quad\xi\in \varGamma_D,\\ 
    \ell(\xi)>0,\quad \text{otherwise}.
\end{array}
\right.\]
For periodic boundary condition, e.g., $\mathcal{G}(v)(\xi)$ is periodic with respect to $\xi$ of the period $P$ in 1D, we can directly substitute trunk input $\xi$ with Fourier basis 
\begin{align*}\{1,\cos(\omega\xi),\sin(\omega\xi),\cos(2\omega\xi),\sin(2\omega\xi),\ldots\}
\end{align*}
with $\omega = 2\pi / P$.

The branch inputs of DeepONet are often high-dimensional. To include the information of input functions, especially for less smooth $v$, more sensors are needed~\cite{lu2021learning}. In some of the examples (Section~\ref{sec: PI-deeponet_results}), we apply principal component analysis (PCA)~\cite{SMAI-JCM_2021__7__121_0} to reduce input dimension. 

\section{Ideal quantum simulation results}\label{sec: ideal_results}

To demonstrate the efficacy of our method, we first use QOrthoNN to approximate certain functions (Section~\ref{sec: func_approx}). Subsequently, we move to the application of quantum DeepONet on learning ODE and PDE problems, including the antiderivative operator (Section~\ref{sec: ODE}), advection equation (Section~\ref{sec: advection}), and Burgers' equation (Section~\ref{sec: burgers}). Finally, we test QPI-DeepONet using the antiderivative operator and Poisson's equation (Section~\ref{sec: PI-deeponet_results}).

We implement the classical training by using the library DeepXDE~\cite{lu2021deepxde}.
After classical training on OrthoNN, we extract the weights and biases and construct a quantum version incorporating quantum layers, applying Qiskit~\cite{Qiskit} for quantum simulation. It is important to note that, in this section, we adopt an idealized scenario during quantum simulation. This approach excludes any quantum and statistical noise, aiming to assess the theoretical accuracy and performance of the quantum model. Table~\ref{tab: quantum-deeponet results} presents the neural network hyperparameters, such as the learning rate and the number of iterations used in Adam optimization~\cite{kingma2014adam}, along with the errors of different examples. The code of all examples is published on GitHub 
(\href{https://github.com/lu-group/quantum-deeponet}{https://github.com/lu-group/quantum-deeponet}). 

\begin{table}[htbp]
\centering
\caption{\label{tab: quantum-deeponet results}\textbf{The default parameters and test $L^2$ relative error for different examples of quantum DeepONet.} For quantum DeepONet, the first number in the ``Depth'' column is the depth of the branch net, and the second number is the depth of the trunk net. The same for the ``Activation'' column. The ``Error'' column presents results for both classical training and ideal simulation, as they are identical for all examples. Here, ``classical'' denotes the use of classical DeepONet to benchmark against the quantum DeepONet. In classical DeepONet, we have utilized a smaller network depth to ensure that the number of trainable parameters remains comparable to those in the quantum version.}
\begin{small}
\begin{tabular}{l|c@{\hspace{7pt}}c@{\hspace{7pt}}c@{\hspace{7pt}}c@{\hspace{7pt}}c|c} \toprule
Example& Depth&Width &Activation &Learning rate&Iteration&Error\\\midrule
 \S\ref{sec: func_approx} Function 1 & 3& 3&Tanh &0.0001&$5\times 10^4$&$0.15\%$\\
 \S\ref{sec: func_approx} Function 2& 4& 10&ReLU &0.0005&$4\times 10^4$&$1.49\%$\\
\S\ref{sec: ODE} Antiderivative $(l=1.0)$& [2,2]&10&ReLU, ReLU &0.001&$3\times 10^4$& $0.49\%$\\
\S\ref{sec: ODE} Antiderivative $(l=0.5)$& [2,2]& 20&ReLU, ReLU &0.001&$3\times 10^4$&$0.84\%$\\
\S\ref{sec: advection} Advection&  [7,7]&21&SiLU, SiLU&0.0005 &$4\times 10^4$& $2.25\%$\\
\S\ref{sec: advection} Advection (classical) &  \textcolor{black}{[4,4]}& \textcolor{black}{21}& \textcolor{black}{SiLU, SiLU} & \textcolor{black}{0.0005} & \textcolor{black}{$4\times 10^4$} & \textcolor{black}{$1.91\%$}\\
\S\ref{sec: burgers} Burgers'&  [6,6]&20&SiLU, SiLU &0.0005&$3\times 10^4$& $1.38\%$\\ 
\textcolor{black}{\S\ref{sec: burgers} Burgers' (classical)}&  \textcolor{black}{[3,3]}& \textcolor{black}{20}& \textcolor{black}{SiLU, SiLU} & \textcolor{black}{0.0005}& \textcolor{black}{$3\times 10^4$}& \textcolor{black}{$1.05\%$}\\ \bottomrule

 \end{tabular}
\end{small}
\end{table}

\subsection{Function approximation\label{sec: func_approx}}
In this section, we adopt two functions to test the accuracy of QOrthoNN. We first consider a function 
\begin{align*}
\text{Function 1:} \qquad f(x) = \frac{1}{1+25x^2}, \quad x \in [-1,1],
\end{align*}
and approximate it using OrthoNN. We choose 80 points for training and 100 points for testing, where $x$ is uniformly sampled in $[-1,1]$. Specifically, for this example, we use the tanh activation function to circumvent the ``dying ReLU'' problem~\cite{lu2019dying},  which is particularly relevant here given the small width of the network. 

\begin{figure}[htbp]
\centering
\includegraphics[width=0.8\linewidth]{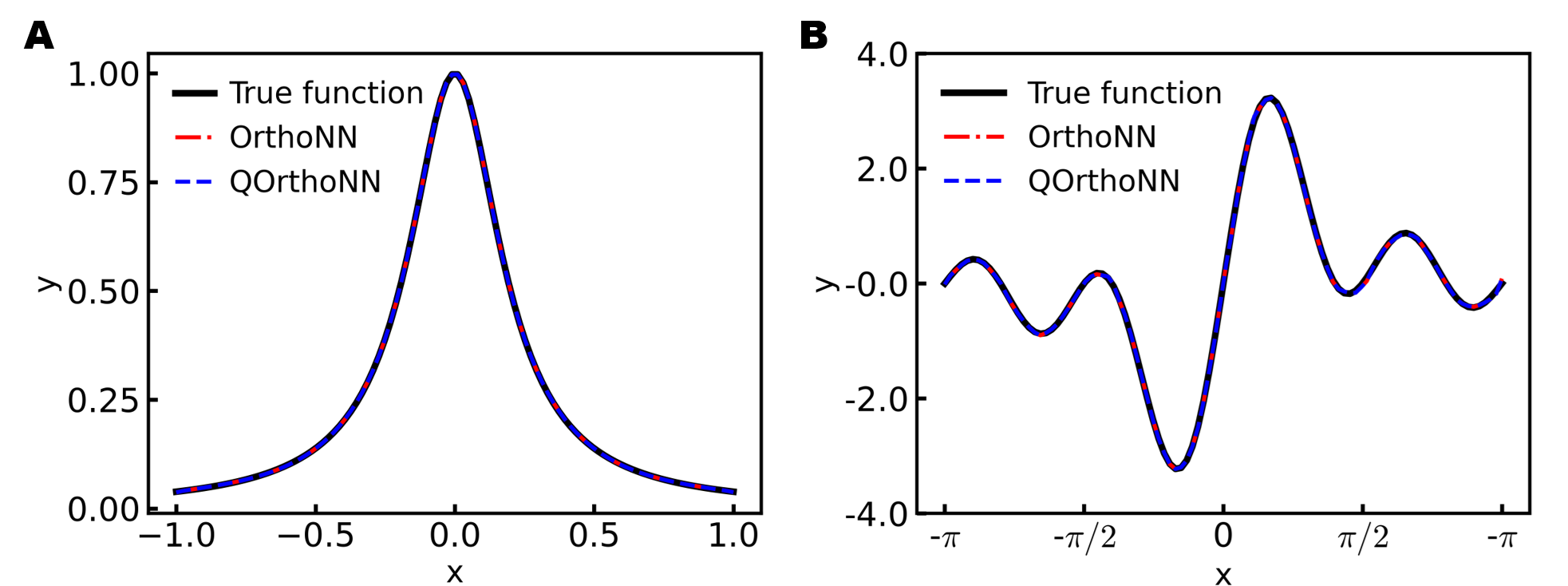}
\caption{\label{fig: func_approx}\textbf{Quantum simulation result of function predictions.} The black, red, and blue lines represent the reference solution, classical prediction of OrthoNN, and ideal quantum simulation result of QOrthoNN, respectively. (\textbf{A}) Results for $f(x)=1/(1+25x^2)$. (\textbf{B}) Results for $f(x)=\sum_{k=1}^{4}\sin(kx)$.}
\end{figure}

For this function 1, we can achieve a small $L^2$ relative error of $0.149\%$ for the testing set after training classically. Following classical training, we construct QOrthoNN using the pyramidal quantum circuit, based on the classically training parameter. The ideal quantum simulation yields an error identical to classical training: $0.149\%$. Essentially, OrthoNN and QOrthoNN are the same neural network, differing only in their prediction methods\textemdash one is executed on a classical computer, while the other is run on a quantum simulator. The results for true function, OrthoNN, and QOrthoNN are plotted in Fig.~\ref{fig: func_approx}A, where the three lines align closely with each other.  

Then, we consider a more complex case for function approximation:
\begin{align*}
\text{Function 2:} \qquad f(x)=\sum_{k=1}^{4}\sin(kx),\quad x\in[-\pi,\pi].
\end{align*}
We use 200 training points, and 100 testing points. Three quantum layers and an output layer with a width of 10 are adopted in the training. The testing error reaches a low relative error of $1.49\%$. The ideal quantum simulation result is also $1.49\%$, highlighting the proficiency of OrthoNN and QOrthoNN (Fig.~\ref{fig: func_approx}B).

\subsection{Antiderivative operator\label{sec: ODE}}
Next, we examine quantum DeepONet by starting with an antiderivate operator:
\begin{equation}
    \frac{du(x)}{dx}=v(x),\quad x\in[0,1],
    \label{equ: ODE}
\end{equation} with initial condition $u(0)=0$. Here, our goal is to learn the operator 
\begin{align*}
\mathcal{G}:v\rightarrow u.
\end{align*}
To generate the input function $v(x)$, we use Gaussian Random Field (GRF):
\begin{align*}
v\thicksim\mathcal{G}\mathcal{P}(0,k_l(x_1,x_2)),
\end{align*}
where $k_l(x_i,x_j)=\exp\left(-\frac{d(x_i,x_j)^2}{2l^2}\right)$ denotes the radial basis function (RBF) kernel. In this context, $d(\cdot,\cdot)$ is the Euclidean distance between two points, and $l$ represents the length scale of the kernel, which modulates the smoothness of the generated function. Specifically, an increase of the value of $l$ leads to a smoother generated function. Therefore, we can adjust $l$ depending on our desired level of function's complexity.

In this example, we explore two scenarios with different length scales: $l=1.0$ and $l=0.5$, corresponding to different size of the network during training. We achieved small errors of $0.49\%$ and $0.84\%$, respectively in these two scenarios for both the training of quantum DeepONet and ideal simulation (Table~\ref{tab: quantum-deeponet results}).

\subsection{Advection Equation\label{sec: advection}}
\begin{figure}[htbp]
\centering
\includegraphics[width=\linewidth]{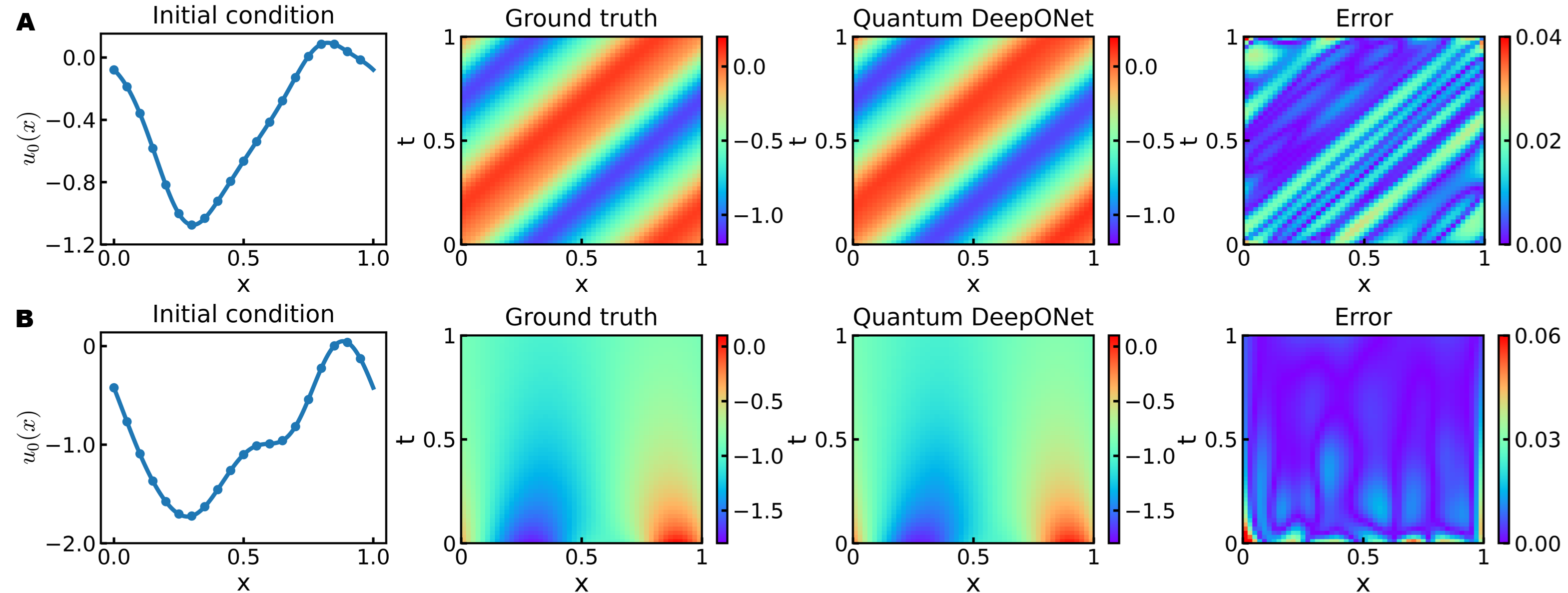}
\caption{\label{fig: advection_burgers}\textbf{Examples of quantum DeepONet prediction for two PDEs.} (\textbf{A}) Advection equation. (\textbf{B}) Burgers' equation.}
\end{figure}

Consider the 1D advection equation:
\begin{align*}
\frac{\partial u}{\partial t}+\frac{\partial u}{\partial x}=0,\quad  x\in[0,1],~t\in[0,1],
\end{align*}
with initial condition $u(x, 0) = u_0(x)$ and periodic boundary condition. Our objective is to learn the operator that maps $u_0(x)$ to the solution $u(x,t)$:
\begin{align*}
\mathcal{G}:u_0(x)\mapsto u(x,t).
\end{align*}
The initial condition $u_0(x)$ is sampled from GRF with Exp-Sine-Squared kernel, formulated as 
\begin{align*}
k(x_i,x_j)=\exp{\left(-\frac{2\sin^2{(\pi d(x_i,x_j)/p)}}{l^2}\right)}.
\end{align*}
Here, $p$ is the periodicity of the kernel and is set to $1$. We choose $l=1.5$ and derive the ground truth using the analytical solution $u(x,t)=u_0(x-t)$. For branch inputs $u_0(x)$, 20 sensors are uniformly placed (see one example in Fig.~\ref{fig: advection_burgers}A left). Regarding the trunk inputs, we employ a grid of $50\times50$ points, covering the range of $x$ and $t$. We implement the ResNet~\cite{he2016deep} architecture in both branch and trunk nets, which has a formulation of $\vect{x}' = \sigma(\vect{W}\vect{x}+\vect{b})+\vect{x}$ for each layer. This approach effectively mitigate the issue of gradient vanishing during training. The final test error of classical prediction reaches $2.25\%$. Ideal simulation of quantum DeepONet yields the same error: $2.25\%$. The quantum DeepONet, with $3081$ trainable parameters, matches the accuracy of the classical DeepONet, which achieves an error of $1.91\%$ using $3171$ trainable parameters (Table~\ref{tab: quantum-deeponet results}). Fig.~\ref{fig: advection_burgers}A provides an example of illustrating the ground truth, predictions of quantum DeepONet, and the absolute error between them. 

\subsection{Burgers' Equation\label{sec: burgers}}
Based on the linear advection equation example, we further examine the non-linear 1D Burgers' equation 
\begin{align*}
\frac{\partial u}{\partial t}+u\frac{\partial u}{\partial x}=\nu\frac{\partial^2 u}{\partial x^2},\quad  x\in[0,1],~t\in[0,1],
\end{align*}
with initial condition $u_0(x)$ and periodic boundary condition, where $\nu=0.05$ is the viscosity. We aim to learn the mapping from $u_0(x)$ to the solution $u(x,t)$. Additionally, recognizing the periodic nature of the output function $u(x,t)$, instead of directly input $\xi$ for trunk net, we expand it to $[\xi,\cos(2\pi\xi),\sin(2\pi\xi),\cos(4\pi\xi),\sin(4\pi\xi)]$. Other neural network settings are the same as Section~\ref{sec: advection}, except for the depth and width. With comparable number of trainable parameters ($2429$ for quantum DeepONet and $2260$ for classical DeepONet), the classical DeepONet achieves an error of $1.05\%$, whereas the quantum DeepONet sustains comparable performance, exhibiting a classical test relative error of $1.38\%$, and an ideal quantum simulation test error of $1.38\%$. An example of the ground truth and prediction of quantum DeepONet is shown in Fig.~\ref{fig: advection_burgers}B.

\subsection{Quantum physics-informed DeepONet \label{sec: PI-deeponet_results}}

In this section, we further show that our quantum DeepONet can also be trained using physics-informed loss term (Eq.~\eqref{eq: physics loss}), without labeled data. We choose the antiderivative equation in Eq.~\eqref{equ: ODE} for comparison with a data-driven case in Section~\ref{sec: ODE}. Additionally, 1D Poisson's equation 
\begin{align*}
\frac{\partial^2 u}{\partial x^2}=v(x),
\end{align*}
with zero Dirichlet boundary condition is also considered for demonstration. The branch inputs in both cases are the $v(x)$ in equations, which are generated by GRF with RBF kernel. To facilitate training and keep PDE residual within a reasonable range, in Poisson's equation, we multiply the generated GRF with a factor of 10 and take the enlarged function as an input sample. The boundary condition is hard constrained using corresponding neural network architecture as mentioned in Section~\ref{sec: QPI-DeepONet}. Zero coordinate shift algorithm~\cite{leng2024zero} is utilized to reduce GPU memory consumption and training time. During the training, the number of input samples is 10000 with batch size = 2000. Adam optimization is used with $2\times 10^5$ iteration. The PDE residual is evaluated at 100 uniformly distributed points in $[0,1]$. During the training, we employed PCA with the original dimension 100. The training results are shown in Table.~\ref{tab: PI-deeponet results}. Ideal quantum simulation results also agree well with classical training results for all of these examples, shown ``$L^2$ relative error'' column.  

 We also conducted experiments without any dimension reduction techniques. For antiderivative with initial condition $l=1$, using branch and trunk net with a depth of 3 and width of 10, the test $L^2$ relative error is $4.05\%$. For comparison, using the same hyperparameters with PCA, which projects the original 100 dimensions down to 10, resulted in an error of $0.76\%$. We believe this difference is due to the critical dependency of QPI-DeepONet on the sampling of input sensors. As derivatives are taken with respect to the inputs, QPI-DeepONet is more sensitive to the input data. PCA enables us to incorporate more information within a limited input dimension. The limitations of current quantum devices compel us to use narrower neural networks, leading to sparse sampling of the branch input. 

\begin{table}[htbp]
\centering
\caption{\label{tab: PI-deeponet results} \textbf{Hyperparameters and training results of QPI-DeepONet for two PDEs with various input function complexity.} The activation function used for all examples is Tanh. The ``$L^2$ relative error'' column presents results for both classical training and ideal simulation, as they are identical for all examples.}
\begin{tabular}{l|ccc|c} \toprule
Example& Number of PCs & Depth& Width & $L^2$ relative error\\\bottomrule
Antiderivative ($l=1$) & 10 & [3,3] & 20&$0.76\%$\\
Antiderivative ($l=0.5$) & 10 & [4,4] & 20&$1.21\%$\\
Antiderivative ($l=0.2$) & 19& [5,5] & 20& $1.91\%$\\
Poisson's ($l=1$)& 10& [3,3] & 20&$0.95\%$\\
Poisson's ($l=0.5$)& 10& [5,5] & 20& $1.55\%$\\
Poisson's ($l=0.2$)& 19& [7,7] &20& $2.31\%$\\ \bottomrule
 \end{tabular}
\end{table}
\section{Effects of noise\label{sec: noise_result}}

Quantum noise is a major obstacle to the practicality of a quantum algorithm in the noisy intermediate-scale quantum (NISQ) era. It emerges from various sources, including the imperfect implementation of quantum operators, undesired environmental or qubit interactions, and erroneous state preparation or measurement. 
During the execution of a quantum circuit, the accumulated errors produced by the noise can destroy any information we intend to obtain. 
Meanwhile, even in a fault-tolerant scenario, the inaccuracy resulting from the finite number of measurements still affects the error level and complexity of the neural network, making it unavoidable to discuss the feasibility of our work on near-term quantum computers. Thus, in Sections~\ref{sec: finite-sampling} and~\ref{sec: depoRBS}, we first provide a theoretical analysis of the effects of finite-sampling noise on the single RBS gate and tomography outputs and a well-known noise channel, depolarizing noise, respectively. 
Then, we demonstrate our noisy simulation results of quantum DeepONet under both types of noise, as well as a more comprehensive noise model emulating a real IBM quantum computer in Section~\ref{sec: noise results}.

\subsection{Finite-sampling noise in tomography}\label{sec: finite-sampling}

In the tomography step (Section~\ref{sec: tomography}), the probabilities ${\rm Pr}[0,\vect{e}_{j}]$ and ${\rm Pr}[1,\vect{e}_{j}]$, for $j\in \{1,...,r\}$, are estimated from the frequencies of measurement outcomes, where $n$ is the input vector dimension, $m$ is the output vector dimension, and $r = \max(m,n)$. This introduces an additional error in the estimation of output vector $\vect{y}$, caused by the finite number of measurements (shots). Let $q^{(0,j)}$ be the probability of measuring $\ket{0,\vect{e}_{j}}$ in tomography layer, and $\hat{q}^{(0,j)}$ be the estimated value from $N_{shot}$ shots. To estimate the size of the finite-sampling error, we first calculate the standard deviation of $\hat{q}^{(0,j)}$.

Let $Z^{(0,j)}_k$ be a Bernoulli random variable
\begin{align*}
    Z^{(0,j)}_k = \begin{cases}
        1 &\text{, if measures $\ket{0,\vect{e}_{j}}$ in $k^{th}$ shot with probability $q^{(0,j)}$} \\
        0 &\text{, otherwise}
    \end{cases}
\end{align*}
and $S^{(0,j)} = \sum_{k = 1}^{N_{shot}} Z^{(0,j)}_k \sim \mathrm{Bin}\left(N_{shot},  q^{(0,j)}\right)$. Thus, we have
\begin{align*}
    \hat{q}^{(0,j)} = \frac{S^{(0,j)}}{N_{shot}}.
\end{align*}
It follows that $\mathrm{E}\left[ \hat{q}^{(0,j)} \right] = q^{(0,j)}$ and, asymptotically, the delta method yields
\begin{align*}
    \left( \sqrt{\hat{q}^{(0,j)}} - \sqrt{ q^{(0,j)} } \right) \xrightarrow{d} \mathcal{N}\left(0,  \frac{1 - q^{(0,j)}}{4N_{shot}} \right),
\end{align*}
where $\cdot  \xrightarrow{d} \cdot $ indicates converge in distribution.
With Eq.~\eqref{equ: tomography}, for $y_j \geq 0$, the standard deviation of the estimated $y_j$ is
\begin{align}
    \mathrm{Std}\left[y_j\right] &= \mathrm{Std}\left[ {\rm sign}(y_j)  \left(2 \sqrt{{\rm Pr}[0,\vect{e}_{j}]}  - \frac{1}{\sqrt{r}} \right) \right] \nonumber \\
    &= 2 \cdot  \mathrm{Std}\left[ \sqrt{\hat{q}^{(0,j)}} \right] \approx  \frac{\sqrt{1 - q^{(0,j)}}}{\sqrt{N_{shot}}} \propto \frac{1}{\sqrt{N_{shot}}},
     \label{eq: finite-sample}
\end{align}
where $q^{(0,j)} \in [0,1]$. 
Similarly, defining $\hat{q}^{(1,j)}$ as the probability of measuring $\ket{1,\vect{e}_{j}}$ in tomography layer and letting $\hat{q}^{(1,j)}$ be its estimate from $N_{shot}$ shots, we also have $\mathrm{Std}\left[y_j\right] \propto 1/N^{-0.5}_{shot}$ for $y_j < 0$.
In conclusion, the finite-sampling error on the estimation of output vector $\vect{y} \in \mathbb{R}^r$  is proportional to $N_{shot}^{-0.5}$ when $N_{shot}$ is large enough.

\subsection{Depolarizing noise on a RBS gate\label{sec: depoRBS}}

The depolarizing noise is a widely adapted noise channel in analyzing the effects of quantum noise on variational quantum circuits~\cite{wang2021noise, stilck2021limitations, fontana2022non, garcia2024effects}. We look closer at how this type of noise affects a quantum layer composed of RBS gates and how the gate's parameter values influence the level of error induced. Note that the statevector representation becomes insufficient to depict the quantum system under the influence of quantum noise. So, we use the density matrix to represent quantum states. A brief introduction to the definition and computation of the density matrix is provided in Appendix~\ref{app: density-matrix}. 

The specific type of noise in our interest is depolarizing noise.
The $n$-qubit depolarizing channel has the expression~\cite{nielsen2010quantum}
\begin{align}
    \mathcal{E}(\rho) &= (1-\lambda) \rho + \lambda \frac{I^{(2^r)}}{2^r}, \label{eq: dep-general}
\end{align}
where $\rho$ is an arbitrary $r$-qubit density matrix, $\lambda$ is a noise parameter, and $I^{(2^r)}$ is a $2^r$-by-$2^r$ identity matrix. Specifically, in a $2$-qubit case, Eq.~\eqref{eq: dep-general} is equivalent to
\begin{align}
     \mathcal{E}(\rho) &= (1-\lambda) \rho + \lambda \frac{I^{(4)}}{4} \nonumber \\
     &= (1-\lambda) \rho + \frac{\lambda}{16} \sum_{i,j \in [4]} (L_i \otimes L_j) \rho (L_i \otimes L_j),\label{eq: 2qdep}
\end{align}
where $L = \{X,Y,Z,I\}$ is the set of Pauli matrices and the $2$-by-$2$ identity matrix $I$. In other words, the effect of $2$-qubit depolarizing noise means there is $1-15\lambda/16$ chance that the state $\rho$ remains unaffected, and an equal chance that any of the 15 possible $2$-qubit Pauli noise occurs on the state $\rho$. To further illustrate the impact of a noisy RBS gate, we consider a noise model 
where a noiseless RBS gate is first applied to the state $\rho$, followed by a depolarizing channel, resulting in the final state $\rho'$. In particular, we have
\begin{align}
    \rho' &= \mathcal{E}\left(U_{RBS} \rho U_{RBS}^\dagger \right) \nonumber \\
    &= (1-\lambda) \left(U_{RBS} \rho U_{RBS}^\dagger \right) + \frac{\lambda}{16} \sum_{i,j \in [4]} (L_i \otimes L_j) \left(U_{RBS} \rho U_{RBS}^\dagger \right) (L_i \otimes L_j).\label{eq: rhoprime-2q}
\end{align}
However, the difference between $\rho$ and $\rho'$ is not in our interest since only the $2^{nd}$, and the $3^{rd}$ elements of the diagonals of $\rho$ and $\rho'$ contain the relevant information.

To put the discussion in the pyramidal circuit setting, let the normalized input vector be $\vect{x} = ( x_1 \quad x_2)^T \in \mathbb{R}^2$, $x_1^2 + x_2^2 = 1$, and the vector after the linear transformation is $\vect{y} = \vect{W} \vect{x}$.
Let $\vect{y}'$ denote the noisy version of $\vect{y}$ due to the depolarizing noise in a RBS gate, and $\cdot^{\circ 2}$ represent the Hadamard (element-wise) square. Because we only encode the entry values of $\vect{x}$ and $\vect{y}$ on the coefficients of $\ket{01}$ and $\ket{10}$, the vector $\vect{y}^{\circ 2}$ is the vector consists of the $2^{nd}$ and $3^{rd}$ elements of the diagonal of $U_{RBS} \rho U_{RBS}^\dagger$ and $\left(\vect{y}^{\circ 2}\right)'$ is the vector of the $2^{nd}$ and $3^{rd}$ elements of the diagonal of $\rho'$. Define function $\diag: \mathbb{R}^{r \times r} \rightarrow \mathbb{R}^{r}$ that extracts the diagonal of a matrix into a vector. In this case, the density matrix $\rho$ is
\begin{align}
    \rho = 
    \begin{pmatrix}
        0 \\ x_1 \\ x_2 \\ 0
    \end{pmatrix}
    \left( 0 \quad x_1 \quad x_2 \quad 0\right) = 
    \begin{pmatrix}
        0 & 0 & 0 & 0\\
        0 & x_{1}^2 & x_{1} x_2 & 0\\
        0 & x_{1} x_2 & x_2^2 & 0\\
        0 & 0 & 0 & 0
    \end{pmatrix}, \label{eq: rho-2q}
\end{align}
and
\begin{align*}
    \vect{y}^{\circ 2} =  \begin{pmatrix} y_1^2 \\ y_2^2 \end{pmatrix} = 
    \begin{pmatrix} \diag\left(U_{RBS} \rho U_{RBS}^\dagger \right)_2 \\  \diag\left(U_{RBS} \rho U_{RBS}^\dagger \right)_3 \end{pmatrix} =
    \begin{pmatrix}
    \left(x_{1} \cos{\theta} + x_2 \sin{\theta}\right)^2 \\
    \left(-x_{1} \sin{\theta} + x_2 \cos{\theta}\right)^2
    \end{pmatrix}.
\end{align*}
Combining Eqs.~\eqref{eq: rhoprime-2q} and~\eqref{eq: rho-2q}, the noisy output is
\begin{align*}
    \left(\vect{y}^{\circ 2}\right)'  &=  \begin{pmatrix} (y'_1)^2 \\ (y'_2)^2 \end{pmatrix} 
    = \begin{pmatrix} \diag\left(\rho' \right)_2 \\  \diag\left(\rho' \right)_3 \end{pmatrix} \\
    &= \frac{1}{4}
    \begin{pmatrix}
     \lambda \left(-x_{1} \sin{\theta} + x_2 \cos{\theta}\right)^2 - 3 \lambda \left(x_{1} \cos{\theta} + x_2 \sin{\theta}\right)^2 + 4 \left(x_{1} \cos{\theta} + x_2 \sin{\theta}\right)^2 \\
     4\left(-x_{1} \sin{\theta} + x_2 \cos{\theta}\right)^2  - 3 \lambda \left(-x_{1} \sin{\theta} + x_2 \cos{\theta}\right)^2 +  \lambda \left(x_{1} \cos{\theta} + x_2 \sin{\theta}\right)^2 
    \end{pmatrix}.
\end{align*}
Determining the sign of each entry of $\vect{y}$ and $\vect{y}'$ requires an additional tomography step as shown in Fig.~\ref{fig: quantum_layer} middle, which may also introduce noise. For simplicity, we only compute the $L^2$ relative error of $|\vect{y}|$,
\tiny
\begin{align}
    \frac{\| |\vect{y}| - |\vect{y}'|\|_2}{\|\vect{|y|}\|_2} &=\| |\vect{y}| - |\vect{y}'|\|_2 
    = \sqrt{ (|y_1| - |y'_1|)^2 + (|y_2| - |y'_2|)^2 } \nonumber\\
    &= \frac{1}2\sqrt{
    \begin{aligned}
    & \left(\sqrt{\lambda \left(x_{1} \sin{\theta} - x_2 \cos{\theta}\right)^2 - 3 \lambda \left(x_{1} \cos{\theta} + x_2 \sin{\theta}\right)^2 + 4 \left(x_{1} \cos{\theta} + x_2 \sin{\theta}\right)^2} - 2 \left|{x_{1} \cos{\theta} + x_2 \sin{\theta}}\right|\right)^2 \\
    & + \left(\sqrt{- 3 \lambda \left(x_{1} \sin{\theta} - x_2 \cos{\theta}\right)^2 + \lambda \left(x_{1} \cos{\theta} + x_2 \sin{\theta}\right)^2 + 4 \left(x_{1} \sin{\theta} - x_2 \cos{\theta}\right)^2} - 2 \left|{x_{1} \sin{\theta} - x_2 \cos{\theta}}\right|\right)^2
    \end{aligned}}         
    \label{eq: l2err-2q}
\end{align}
\normalsize
since $|\vect{y}|$ is a normalized vector. Eq.~\eqref{eq: l2err-2q} shows that the $L^2$ relative error of $|\vect{y}|$ is a periodic function with respect to $\theta$ in a period of $\pi/2$ and the value of $\lambda$ controls the amplitude of the function. The reverse triangle inequality guarantees that the $L^2$ relative error of $\vect{y}$ is always lower-bounded by that of $|\vect{y}|$
\small
\begin{align*}
     \frac{\| |\vect{y}| - |\vect{y}'|\|_2}{\|\vect{ |y| }\|_2} 
    = \sqrt{ (|y_1| - |y'_1|)^2 + (|y_2| - |y'_2|)^2 }
    \leq \sqrt{ (y_1 - y'_1)^2 + (y_2 - y'_2)^2 }
    = \| \vect{y} - \vect{y}'\|_2  = \frac{ \| \vect{y} - \vect{y}'\|_2}{\|\vect{y}\|_2}.
\end{align*}
\normalsize

\begin{figure}[htb]
    \centering
    \includegraphics[width=0.59\linewidth]{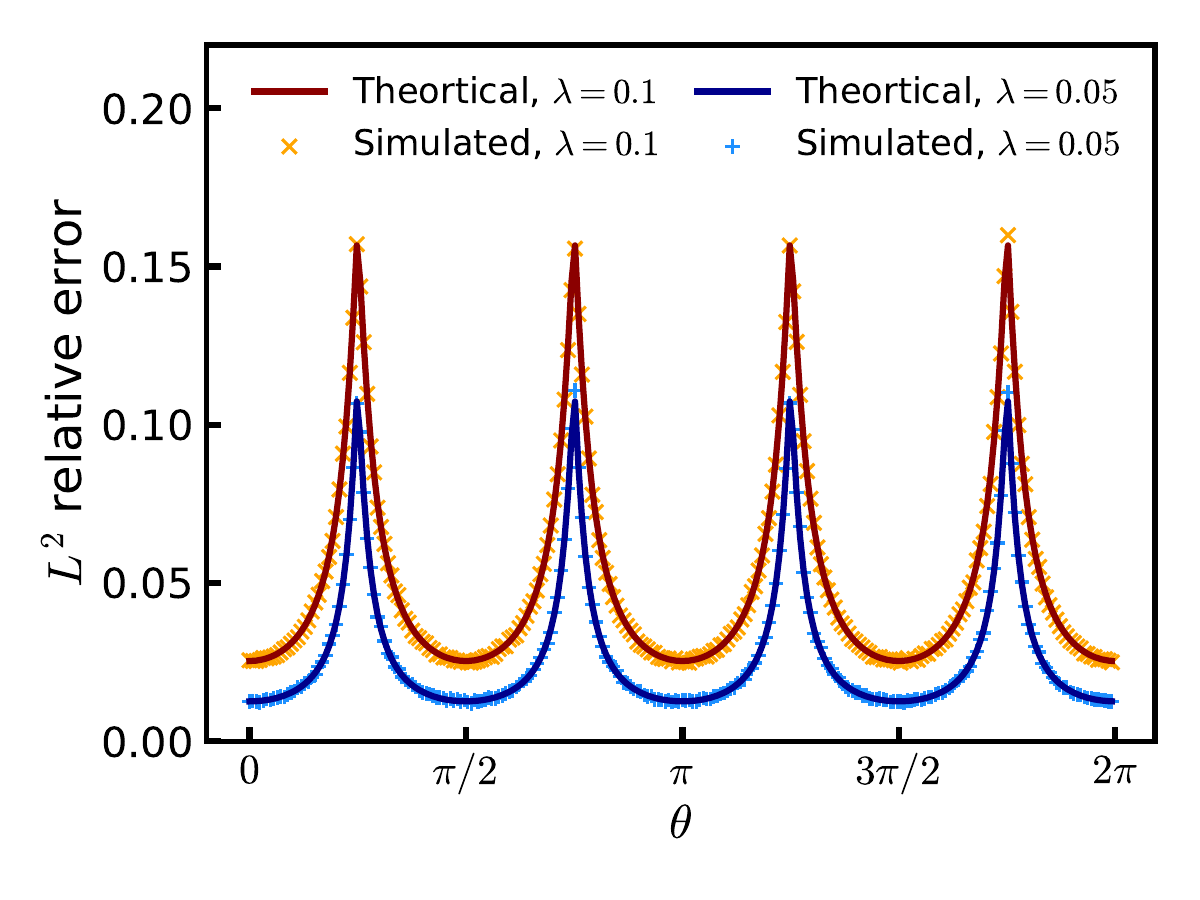}
    \caption{\textbf{Errors of the output vector, $\vect{y}$, due to the 2-qubit depolarizing noise on a single RBS gate as a function of the angle of the RBS gate, $\theta$.} The initial state in the circuit is $\left[ 0,\, 1/\sqrt2, \, 1/\sqrt2, \, 0\right]^T$. Each simulated data point is averaged from 100,000 samples in the Qiskit Aer simulator with a simulated depolarizing noise model.}
    \label{fig: rbs-dep}
\end{figure}

We numerically illustrate Eq.~\eqref{eq: l2err-2q} in Fig.~\ref{fig: rbs-dep} with $x_1 = x_2 = 1/\sqrt2$ for $\lambda = 0.1$ and $\lambda=0.05$. To show that our computation is consistent with the noise model in Qiskit Aer, we also provide the estimations from the samples in the Qiskit Aer simulator with simulated depolarizing noise models. Each data point in the simulated case in Fig.~\ref{fig: rbs-dep} is the average of 100,000 samples from the simulator. Since $\vect{y}$ and $\vect{y}'$ are non-negative in the experiments in Fig.~\ref{fig: rbs-dep}, the plot equivalently shows the $L^2$ relative error of $\vect{y}$. 

\subsection{Noisy simulation results\label{sec: noise results}}

In our subsequent investigation, we adopt two types of noise models to assess the accuracy of quantum layers under noisy conditions. 
The first approach, the simplified noise model, incorporates only 1-qubit and 2-qubit depolarizing noise channels on all basis gates.
The goal of using this model is to examine the noise resilience of pyramidal circuits and the effects of the error mitigation technique on this well-researched noise channel, depolarizing noise channel.
In the experiments, we select several different values for 1-qubit noise parameter $\lambda$, defined in Eq.~\eqref{eq: dep-general}, and set the 2-qubit noise parameter $\lambda':=0.8\lambda$. This is to guarantee both noise channels have the same error rate. Recall Eq.~\eqref{eq: dep-general}, 1-qubit depolarizing noise channel has the expression
\begin{align*}
    \mathcal{E}_{dep}(\rho) = (1-\lambda) \rho + \lambda \frac{I}{2} = \left(1-\frac{3}{4}\lambda\right) \rho + \frac{\lambda}{4} \left(X\rho X + Y\rho Y + Z\rho Z\right).
\end{align*}
So the error-free probability is $1-0.75\lambda$.  Similarly, the error-free probability for a 2-qubit depolarizing noise channel is $1 - 0.9375 \lambda'$, as shown in Eq.~\eqref{eq: 2qdep}. 
By setting $\lambda' = 0.8\lambda$, the two probabilities become equal. In our experiments, we choose the values of $\lambda$ from $0$ to $2 \times 10^{-3}$ since the gate error rates on real IBMQ quantum computers are of a similar scale, as indicated in Table~\ref{tab: real-rates}.

\begin{table}[htbp]
\centering
\caption{\label{tab: real-rates} \textbf{1-qubit basis gate error rates, $0.75\lambda$, among all qubits on selected IBMQ quantum computers (data collected on May 21, 2024~\cite{ibmq}).}}
\begin{tabular}{l|cccc} \toprule
        &ibm\_osaka &ibm\_brisbane &ibm\_sherbrooke &ibm\_torino\\\bottomrule
Average & $1.37 \times 10^{-3}$ &$6.29 \times 10^{-4}$ &  $2.07 \times 10^{-4} $& $1.53\times 10^{-3}$\\
Median & $2.68 \times 10^{-4}$ & $2.38  \times 10^{-4}$ &  $ 5.08 \times 10^{-4}$  & $3.52 \times 10^{-4}$ \\
\bottomrule
 \end{tabular}
\end{table}

While the first approach aims for a direct and intuitive evaluation of the accuracy of pyramidal
circuit under a noisy environment, depolarizing noise is insufficient to fully reflect the noise in real quantum computers and the 2-qubit gates usually have less fidelity than 1-qubit gates~\cite{wright2019benchmarking, georgopoulos2021modeling, amico2023defining}.
To fill this gap, we also carry out experiments with the second approach: the backend-noise model from Qiskit Aer~\cite{Qiskit, dasgupta2024impact}. The backend-noise model is in composite of 
\begin{itemize}
    \item measurement noise: emulated by classical 1-qubit bit-flip error in the measurement;
    \item gate noise: emulated by the combination of 1-qubit depolarizing error and thermal relaxation error, while the 2-qubit error operator is the tensor product of 1-qubit error operators.
\end{itemize}
The parameters of the backend-noise model come from the regular benchmarking tests performed by the device vendor.
By comparing these models, we can identify the feasibility of our quantum neural network and provide benchmarks for the improvement of near-term quantum computers.

\subsubsection{Function approximation}
To demonstrate the impact of quantum noise, we first choose the example of function approximation $f(x)=1/(1+25x^2)$ in Section~\ref{sec: func_approx} function 1. All of the following results are calculated in a Qiskit simulator. 

\begin{figure}[htb]
\centering
\includegraphics[width=0.8\linewidth]{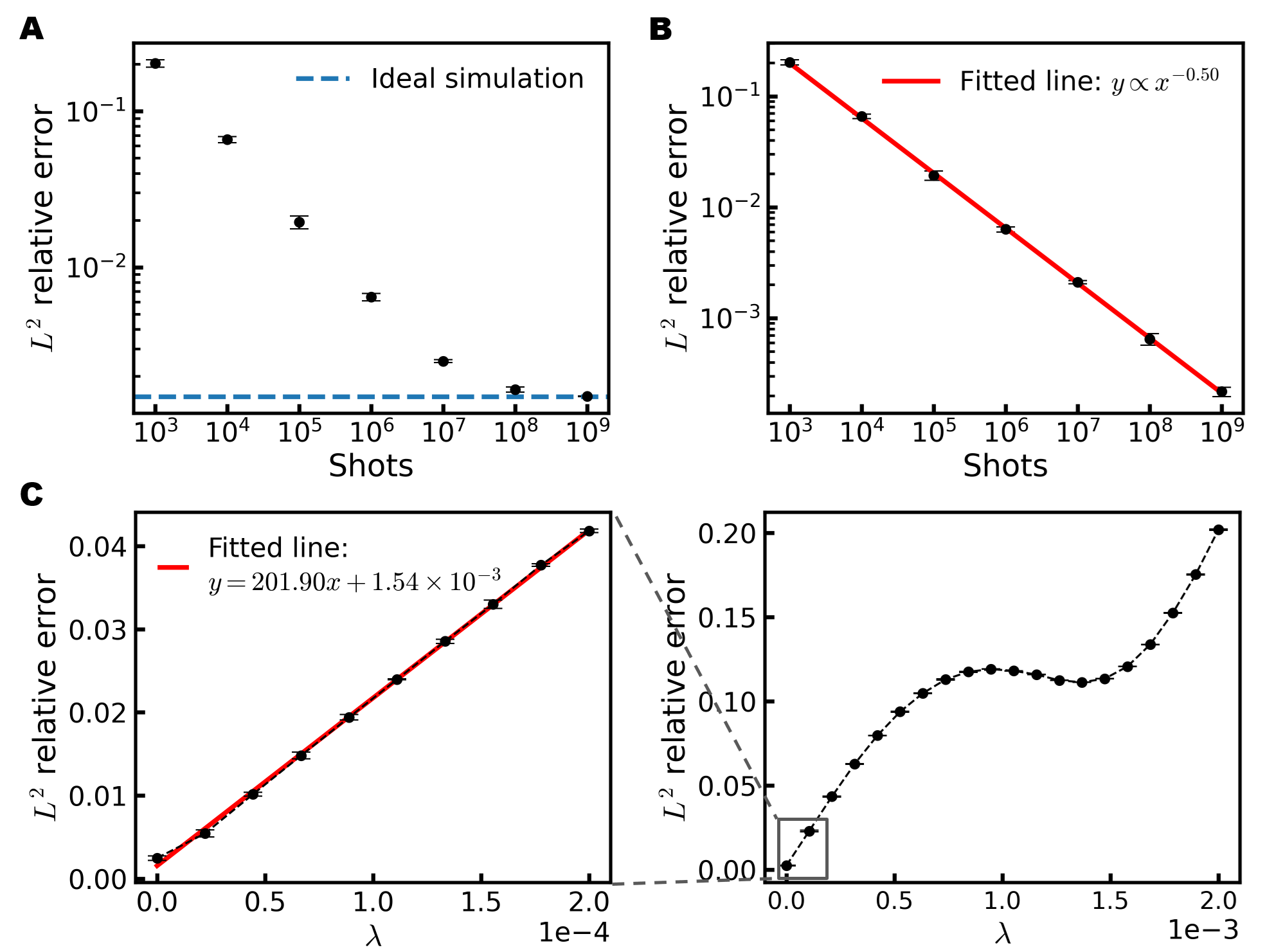}
\caption{\label{fig: func_noise}\textbf{ Effect of quantum noise on function approximation example of $f(x)=1/(1+25x^2)$.} (\textbf{A} and \textbf{B}) Finite-sampling noise at different number of shots. (A) $L^2$ error between shots-based results and true function with different shots, compared with ideal simulation. (B) $L^2$ error between shots-based and ideal simulation. (\textbf{C}) Depolarizing noise model for different depolarizing parameters. In both cases, the number of shots is set to be $10^7$.}
\end{figure}

We investigate the impact of finite-sampling error by varying the number of shots, $N_{shot}$, i.e., how many times we do the measurement to reconstruct the quantum state. The error with respect to true function value decreases when we increase the number of shots (Fig.~\ref{fig: func_noise}A). In this example, when the number of shots reaches $10^8$, the shots-based simulation result is close to the ideal simulation result. We further analyzed the error between shots-based and ideal simulation (Fig.~\ref{fig: func_noise}B), which is exactly the finite-sampling error mentioned in Section~\ref{sec: finite-sampling}. The error in Fig.~\ref{fig: func_noise}B is proportional to $N^{-0.5}_{shot}$, which fits perfectly with the theoretical results in Eq.~\eqref{eq: finite-sample}.

We further include depolarizing error, as introduced in Section~\ref{sec: depoRBS}, to estimate the effect of quantum gate noise. The error mitigation scheme is applied here.  When $\lambda$ is within $[0,2\times 10^{-4}]$,  the error increases almost linearly with $\lambda$ (Fig.~\ref{fig: func_noise}C left). When we expand the range to $[0,2\times 10^{-3}]$, the error increases non-linearly and reaches a plateau at approximately $\lambda=10^{-3}$. As $\lambda$ reaches $2\times 10^{-3}$, the simulated error reaches around $20\%$ (Fig.~\ref{fig: func_noise}C right). 

In order to simulate the performance of our quantum neural network on real quantum computer, we further adapt the backend-noise model in Qiskit. Here, we choose ibm\_brisbane backend, loading the corresponding noise parameters for the simulation. 
The error turns out to be $14.4\%$, suggesting some more sophisticated error mitigation methods will needed in future work. Since the scale of error is already too large in the simplest QOrthoNN experiment, the backend-noise model will not be tested in further experiments.


\subsubsection{Antiderivative operator}\label{sec: noise_ODE}

\begin{figure}[htb!]
\centering
\includegraphics[width=0.8\linewidth]{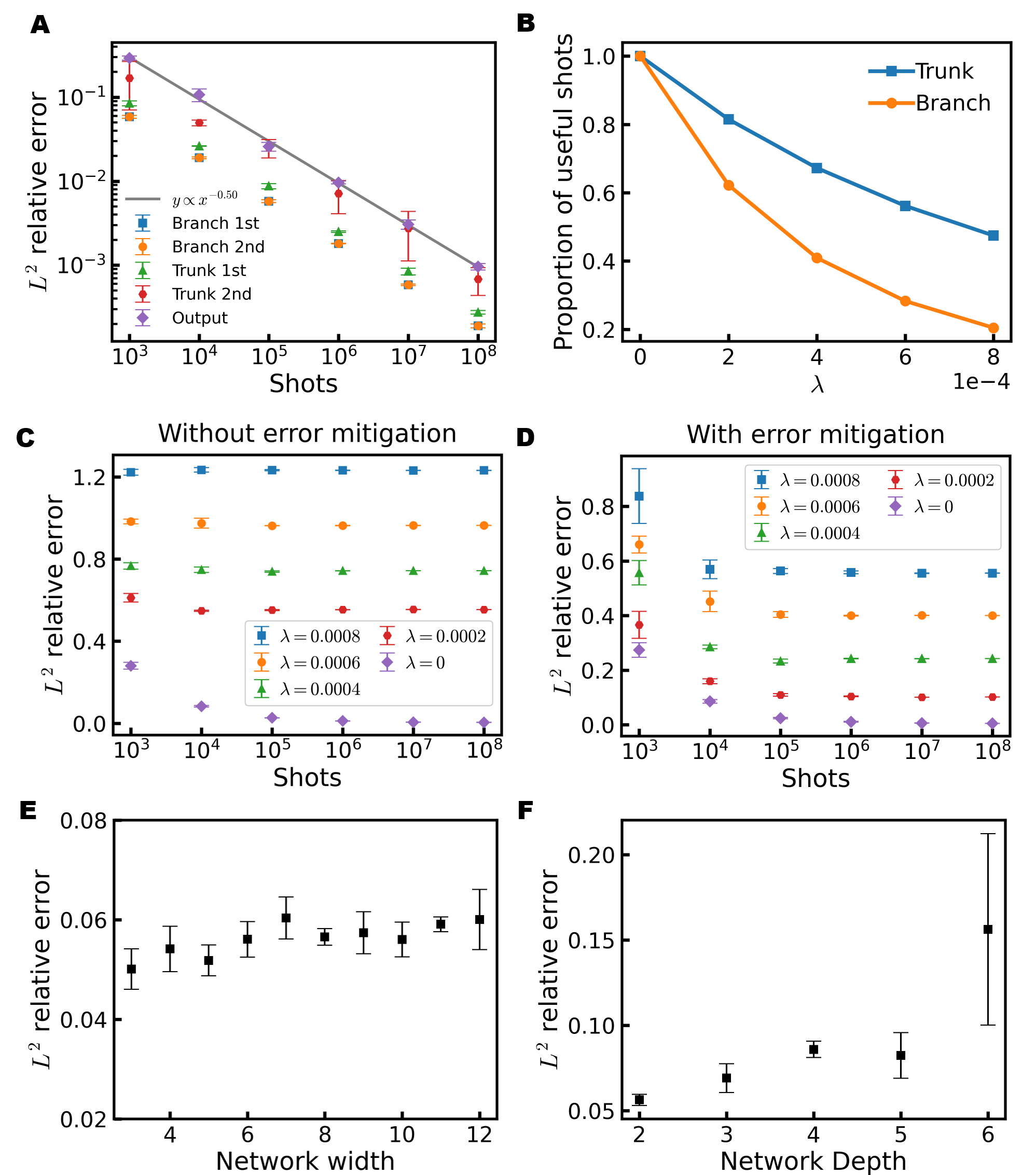}
\caption{\label{fig: ODE_noise}\textbf{Effect of noise on quantum DeepONet for the example of antiderivate operator.} (\textbf{A}) $L^2$ relative error between shots-based and ideal simulation results for different layers. (\textbf{B}) Proportion of useful shots in total shots at different depolarizing noise levels $\lambda$ when implementing error mitigation. (\textbf{C} and \textbf{D}) $L^2$ error between simulation results at different depolarizing level $\lambda$ and true solution. (C) Error mitigation is disabled. (D) Error mitigation is used. (\textbf{E} and \textbf{F}) The error for different neural network size. We set $\lambda = 10^{-4}$ for all gates and fixed the number of shots at $10^7$. For each neural network size, we performed classical training 5 times until the test error is reduced to $3\%$. Each training run is quantumly simulated 3 times. The average and uncertainty of all 15 noisy simulation results were then calculated. (E) The network depth of both the branch and trunk nets is fixed at 5, while the width of both is varied simultaneously. (F) The network width of both the branch and trunk nets is fixed at 10, while the depths of both networks are varied simultaneously.}
\end{figure}

The impact of quantum noise on quantum DeepONet is also investigated using the antiderivate operator example (Section~\ref{sec: ODE}) when $l=1.0$. In Fig.~\ref{fig: ODE_noise}A, we plot the simulated finite-sampling error the scale of which is proportional to $N^{-0.5}_{shot}$, as expected in Eq.~\eqref{eq: finite-sample}. When depolarizing quantum noise is considered, the error mitigation method is discussed in Section~\ref{sec: tomography} can be applied. Although error mitigation helps eliminate undesired results caused by quantum noise, it also reduces the number of shots that are ultimately usable. It is obvious that 
\begin{align*}
\text{useful shots} \approx C \times\text{total shots},
\end{align*}
with $C=1.0$ when $\lambda=0$. The parameter $C$ decreases as $\lambda$ increases (Fig.~\ref{fig: ODE_noise}B) because higher levels of noise produce more unreasonable results. 

The post-selection on the measurement outcomes significantly reduces the error in noise cases with both finite-sampling and depolarizing noise, comparing Fig.~\ref{fig: ODE_noise}C with D. In Fig.~\ref{fig: ODE_noise}D, where error mitigation is not applied and both unary and non-unary results are accepted, the error shows little reduction as number of shots increases. 
This is because the finite-sampling error is relatively insignificant under the influence of depolarizing error in this example, as can be seen by comparing the scales of errors in Figs.~\ref{fig: ODE_noise}A and D. Therefore, increasing shots, which only reduce the finite-sampling error, is not effective without error mitigation. However, with error mitigation, the overall error is significantly lower, making the reduction of finite-sampling error more obvious in the plot (Fig.~\ref{fig: ODE_noise}C). 

We also investigated how the network size can affect the error of the noisy model (Figs.~\ref{fig: ODE_noise}E and F). For each neural network size, we performed classical training 5 times. The networks were trained until the test error was reduced to $3\%$. For each trained network, we quantum-simulated 3 times. The parameters of simulations included $10^7$ shots and $\lambda=10^{-4}$ for depolarizing noise. It is important to note that even though the test error remained the same across classical training runs, the noisy simulation results varied. We believe this variation arises because the network converges to different parameter values in each training run, leading to different levels of error due to depolarizing noise. This observation aligns with our discussion in Section~\ref{sec: depoRBS}, which shows the parameters of RBS gates also influence the magnitude of errors. By comparing the two plots, we conclude that the error increases almost exponentially with increasing network depth. In contrast, when only the network width is increased, the error shows minimal growth within our experiment range. Therefore, quantum DeepONet shows resilience to noise with respect to network width. In practice, to minimize quantum noise, it is advisable to opt for wider rather than deeper neural networks.

\section{Conclusions}
\label{sec: conclusion}

We proposed quantum DeepONet, which can be both data-driven and physics-informed. Experimental results were conducted to confirm that quantum DeepONet performs efficiently in solving different PDEs. 
We further considered the impact of different noise models in simulation and benchmarked the noise level and corresponding accuracy. 

There are a few limitations in our current implementation.
Based on the unary encoding, the quantum DeepONet currently could not handle large network width due to the limitation on the number of qubits and connectivities in the existing quantum devices, and the in-effectiveness of simulation on classical computers.
Also, the circuit in the quantum layer still has the $\mathcal{O}(n)$ depth, which is beyond the $\mathcal{O}(\log n$) bound on producing entanglement on NISQ devices~\cite{yan2023limitations}.
On the other hand, although such demand on the number of qubits can be greatly reduced by removing the unary encoding, the increased cost of data loading and data tomography resulting from this change will require further analysis.
 \textcolor{black}{It may require a design to apply non-linear transformation without collapsing the quantum state, such as the method in~\cite{Moreira2023}.}
 \textcolor{black}{In terms of experiment settings} in the noise simulation, both tested noise models do not include coherent noise and non-local noise such as cross-talk. A more complicated and realistic noise model is needed to examine the noise resilience of our design. Additionally, our future work will explore extending quantum DeepONet to accommodate more advanced architectures~\cite{cai2021deepm, lu2022multifidelity}, which will allow us to address a broader range of applications.

\section*{Acknowledgments}

This work was supported by the U.S. Department of Energy Office of Advanced Scientific Computing Research under Grants No.~DE-SC0025592 and No.~DE-SC0025593, and the U.S. National Science Foundation under Grant No.~DMS-2347833. We acknowledge helpful discussions with Yunjia Yang and Min Zhu.

\appendix
\section{Model of quantum computing}\label{sec: qcintro}

Our work utilizes the quantum circuit model. It is an analogy to the classical circuit where a series of gates are conducted to perform computation. For a basic quantum circuit, there are three components: an initial quantum state, a series of quantum gates, and measurements. The initial state stores the initial information, which is then changed by the sequence of quantum gates. After the computation, the state is measured to get classical bits as the final outputs. The unit of quantum information is a qubit, analogizing to a bit in classical information. 

In most of our work, we use statevector representation for quantum states. That is, an $n$-qubit quantum state is a vector in $\mathbb{C}^{2^n}$. Such a quantum state is often written as a linear combination of basis states. For example, a general 1-qubit state $\ket{\psi}$ is
\begin{align*}
    \ket{\psi} = \alpha \ket{0} + \beta \ket{1},
\end{align*}
where the notation $\ket{\cdot}$ represents a statevector, basis state $ \ket{0}$ is $[1 \,\, 0]^T$, basis state $ \ket{1}$ is $[0 \,\, 1]^T$, and $|\alpha|^2 + |\beta|^2 = 1$ for complex numbers $\alpha$ and $\beta$. So, when we measure the state $\ket{\psi}$, there is an $|\alpha|^2$ chance to obtain a classical bit $0$ and $|\beta|^2$ chance to obtain a classical bit $1$. If neither $\alpha$ nor $\beta$ is 0, then the quantum state is in the superposition of state $\ket{0}$ and $\ket{1}$. Similarly, a 2-qubit state is a linear combination of 2-qubit bases and the squared norms of coefficients sum to 1. The 2-qubit basis states are $\ket{00} = \ket{0} \otimes \ket{0}$, $\ket{01} = \ket{0} \otimes \ket{1}$, $\ket{10} = \ket{1} \otimes \ket{0}$, and $\ket{11} = \ket{1} \otimes \ket{1}$, where the operator $\otimes$ is the Kronecker product. If a 2-qubit state cannot be factored into the tensor product of two 1-qubit states, then this 2-qubit state is entangled. 

An $n$-qubit quantum logic gate is a $2^{n}$-by-$2^n$ unitary matrix. 
Several common 1-qubit gates are
\begin{align*}
    X &= \begin{bmatrix} 0 &1 \\ 1 &0 \end{bmatrix},
    Y = \begin{bmatrix} 0 &-i \\ i &0 \end{bmatrix},
    Z = \begin{bmatrix} 1 &0 \\ 0 &-1 \end{bmatrix},
    H = \frac{1}{\sqrt{2}}\begin{bmatrix} 1 &1 \\ 1 &-1 \end{bmatrix},
    S = \begin{bmatrix} 1 &0 \\ 0 &i \end{bmatrix},\\
    R_x(\theta) &= \begin{bmatrix} \cos\frac{\theta}{2} &-i\sin\frac{\theta}{2} \\ -i\sin\frac{\theta}{2} &\cos\frac{\theta}{2} \end{bmatrix}, 
    R_y(\theta) = \begin{bmatrix} \cos\frac{\theta}{2} &-\sin\frac{\theta}{2} \\ \sin\frac{\theta}{2} &\cos\frac{\theta}{2} \end{bmatrix}, 
    R_z = \begin{bmatrix} e^{-i\theta/2} &0 \\ 0 &e^{i\theta/2} \end{bmatrix},
\end{align*}
and two widely used 2-qubit controlled gates are
\begin{align*}
    CNOT = \ket{0}\bra{0}\otimes I + \ket{1}\bra{1}\otimes  X\text{ and }
    CZ = \ket{0}\bra{0}\otimes I + \ket{1}\bra{1}\otimes  Z,
\end{align*}
where $I$ is the 2-by-2 identity matrix.

\begin{figure}[htb]
    \centering
    \includegraphics[width=0.3\linewidth]{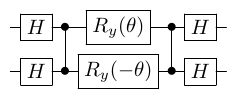}
    \caption{An implementation of $U_{RBS}(\theta)$ according to Ref.~\cite{Landman2022quantummethods}, where the symbol of two connected dots between $H$ and $R_y$ gates is the $CZ$ gate.}
    \label{fig: rbs-impl}
\end{figure}

If an 1-qubit gate $U$ applies on the first qubit of a 2-qubit state $\ket{\chi}$ and another 1-qubit gate $V$ applies on the second qubit simultaneously, the resultant computation is $(U \otimes V)\ket{\chi}$. Based on the gate definitions introduced above, an implementation of $U_{RBS}(\theta)$ according to~\cite{Landman2022quantummethods} is shown in Fig.~\ref{fig: rbs-impl}. It can be verified that
\begin{align*}
    U_{RBS}(\theta) = [H \otimes H]CZ[R_y(\theta) \otimes R_y(-\theta) ] CZ [H \otimes H].
\end{align*}

\section{Quantum states in the density-matrix representation\label{app: density-matrix}}

A density matrix represents a quantum state in quantum information, providing a more general description than the statevector. In quantum computing, density matrices often come when the discussion includes quantum noise because quantum noise, such as the depolarizing channel in Section~\ref{sec: depoRBS}, can result in non-unitary evolution. The resultant quantum system may have $p_k$ probability in the state $\ket{\psi_k}$ for multiple different indices $k$, making a single statevector insufficient to depict it. To express this system in a density matrix $\rho$, we have
\begin{align*}
    \rho = \sum_{k} p_k \ket{\psi_k} \bra{\psi_k},
\end{align*}
where $\sum_k p_k = 1$. Thus, the density matrix $\rho$ is trace-one, Hermitian, and positive semidefinite~\cite{nielsen2010quantum}. The state evolution governed by the unitary operator $U$ is computed by
\begin{align*}
    \rho \overset{U}{\rightarrow} U \rho U^\dagger.
\end{align*}
The non-unitary evolution can be described similarly to the depolarizing noise channel Section~\ref{sec: depoRBS}.




\bibliographystyle{quantum}
\bibliography{main}

\end{document}